%% file: basesafe.tex
\documentclass[sigconf,authorversion]{acmart}

\setcopyright{acmcopyright}
\copyrightyear{2020}
\acmYear{2020}
\setcopyright{acmlicensed}\acmConference[WiSec '20]{13th ACM Conference on Security and Privacy in Wireless and Mobile Networks}{July 8--10, 2020}{Linz (Virtual Event), Austria}
\acmBooktitle{13th ACM Conference on Security and Privacy in Wireless and Mobile Networks (WiSec '20), July 8--10, 2020, Linz (Virtual Event), Austria}
\acmPrice{15.00}
\acmDOI{10.1145/3395351.3399360}
\acmISBN{978-1-4503-8006-5/20/07}

\input{packages.tex}

\begin{document}

%\title{Rehosting and Fuzzing of Mobile Basebands}
\title{BaseSAFE: Baseband SAnitized Fuzzing through Emulation}

\author{Dominik Maier}
\email{dmaier@sect.tu-berlin.de}
\affiliation{%
  \institution{TU Berlin}
  %\streetaddress{P.O. Box 1212}
}
\author{Lukas Seidel}
\email{seidel.1@campus.tu-berlin.de}
\affiliation{%
  \institution{TU Berlin}
  %\streetaddress{P.O. Box 1212}
}
\author{Shinjo Park}
\email{pshinjo@sect.tu-berlin.de}
\orcid{0000-0002-8569-1327}
%\email{webmaster@marysville-ohio.com}
\affiliation{%
  \institution{TU Berlin}
  %\streetaddress{P.O. Box 1212}
}
% \affiliation{%

\renewcommand{\shortauthors}{Maier~et al.}

\input{abstract.tex}




\maketitle

\input{introduction.tex}

\input{background.tex}

\input{related.tex}

\input{rehosting.tex}

\input{evaluation.tex}

\input{conclusion.tex}

%\begin{acks}
%\end{acks}

\bibliographystyle{ACM-Reference-Format}
\bibliography{basesafe}

%\appendix

\end{document}

%% file: packages.tex
\newif\ifclean
\cleantrue
\cleanfalse

\usepackage{adjustbox}
\usepackage{array}
\usepackage{eurosym}
\usepackage{amsmath,amsfonts}
\usepackage{xspace}
\usepackage{color}
\usepackage{msc}
\usepackage{calc}
\usepackage{multirow}
\usepackage{epsfig}
\usepackage{graphicx}
\graphicspath{{./figures/}}
\usepackage{boldline}
\usepackage{pifont}
\usepackage{multirow}
\usepackage{rotating}
\usepackage{footnote}
\usepackage{hyperref}
\usepackage{cleveref}
\usepackage{url}

\usepackage{tabularx}
\newcolumntype{Y}{>{\centering\arraybackslash}X}
\usepackage{tabu}
\usepackage{environ}
\usepackage{makecell}
\usepackage{subcaption}
\usepackage{threeparttable}
\usepackage{adjustbox}

\usepackage{pdfcomment}

\usepackage{soul}

\usepackage{relsize}

\usepackage{listings, listings-rust}

\def\dash---{\kern.16667em---\penalty\exhyphenpenalty\hskip.16667em\relax}

\newcommand\AFL{AFL\nolinebreak[4]\hspace{-.05em}\raisebox{.45ex}{\relsize{-2}{\textbf{++}}}}

\newcommand\Cpp{C\nolinebreak[4]\hspace{-.05em}\raisebox{.40ex}{\relsize{-2}{\textbf{++}}}}

%% file: abstract.tex
\begin{abstract}
Rogue base stations are an effective attack vector.
Cellular basebands represent a critical part of the smartphone's security: they parse large amounts of data even before authentication.
They can, therefore, grant an attacker a very stealthy way to gather information about calls placed and even to escalate to the main operating system, over-the-air.
In this paper, we discuss a novel cellular fuzzing framework that aims to help security researchers find critical bugs in cellular basebands and similar embedded systems.
BaseSAFE allows partial rehosting of cellular basebands for fast instrumented fuzzing off-device, even for closed-source firmware blobs.
BaseSAFE's sanitizing drop-in allocator, enables spotting heap-based buffer-overflows quickly.
Using our proof-of-concept harness, we fuzzed various parsers of the Nucleus RTOS-based MediaTek cellular baseband that are accessible from rogue base stations.
The emulator instrumentation is highly optimized, reaching hundreds of executions per second on each core for our complex test case, around 15k test-cases per second in total.
Furthermore, we discuss attack vectors for baseband modems.
To the best of our knowledge, this is the first use of emulation-based fuzzing for security testing of commercial cellular basebands.
Most of the tooling and approaches of BaseSAFE are also applicable for other low-level kernels and firmware.
Using BaseSAFE, we were able to find memory corruptions including heap out-of-bounds writes using our proof-of-concept fuzzing harness in the MediaTek cellular baseband.
BaseSAFE, the harness, and a large collection of LTE signaling message test cases will be released open-source upon publication of this paper.
\end{abstract}

%% file: introduction.tex
\section{Introduction}
Attacks on mobile basebands are possible from adjacent base stations, which are built to work over fairly large distances.
The targets move constantly, carrying their phone with them, so a rogue base station can potentially attack a large number of phones every day.
Even though operating systems running on the phone's basebands are large attack vectors, parsing many different types of signaling messages, even prior to authentication, the systems are obscure and little research exists.
Each larger chipset vendor for smartphones develops and ships their own stack.
Any sort of automated analysis, if it exists at all, is kept locked away behind the vendor's doors, together with documentation and specifications of their systems.
This gave us the reason to take a closer look at one of them.
Since the systems are written in C/\Cpp{}, known to be haunted by memory corruptions, we set out to build a usable open-source fuzzing environment, BaseSAFE.
As the initial target, we chose MediaTek, one of the large vendors.
Their chips are common in many sought-after mid-tier phones like the \emph{Xiaomi Redmi Note 8 Pro}, with offerings from most vendors, including Motorola, Nokia, HTC, and others.
The cellular baseband has close ties to the mobile operating system.
Calls and data are routed through it and most of the lower-level interactions,
such as establishing a call and selecting a base station, are done directly in 
the baseband but need to be displayed to the user.

%TODO: till which extent shall we mention the previous baseband attacks?
Baseband firmware has one of the widest attack vectors of all components in modern smartphones.
Every cellular network usage passes through the baseband.
Countless high-complexity protocol parsers are part of the firmware.
%, reacting to arriving data in internal IPC-queues and featuring high-complexity protocol parsers.
However, to this date, security testing of mobile basebands is either performed in a black box fashion or through manual static analysis.
Baseband and device manufacturers are trying to limit the direct access to the baseband in various ways, such as blocking JTAG access as well as encrypting parts of or the entire baseband firmware~\cite{Golde2016, Miru2017}.
The secrecy and closed-sourceness, sometimes scrambled firmware, and usages of relatively unknown architectures, such as Qualcomm Hexagon, increase the barrier to mobile baseband analysis, while at the same time making it a more interesting and rewarding target.
%Baseband security analysis on the binary blobs is usually done 
%manually through time-consuming reverse engineering---often supported by source 
%code leaks~\cite{Miru2017}. 
%To the best of our knowledge, open automated testing on baseband firmware itself is still in a developing stage. 

In this paper we take a novel approach: by rehosting parts of a memory dump of a major cellular baseband, we are able to run main event handlers in our analysis platform.
We propose \emph{BaseSAFE}, a platform combining speedy emulation with fuzzing and heap sanitization.
Building on the popular \emph{Unicorn engine} emulator, BaseSAFE allows us to 
perform coverage-guided fuzzing on the MediaTek baseband.
We automatically map signaling messages to their respective functions using coverage feedback, to keep false-positives low.
BaseSAFE implements a custom drop-in sanitizing heap allocator for Unicorn, which can replace any baseband-internal allocation mechanism to uncover heap corruptions and use-after-frees with fuzzing.

The sample harness of BaseSAFE for LTE Radio Resource Control (RRC)~\cite{3gpp.36.331} messages ships with thousands of unique, minimized, valid inputs for a layer 3 parser in the MediaTek baseband, the \texttt{errc\_event\_handler\_main} function of the firmware.
All samples trigger different code paths in the parser and will likely be a good set to fuzz the same parsers in other baseband firmwares.
The signaling messages can be sent to the phone by a modified base station unauthenticated.

%\subsection*{Contributions}
The key contributions of this paper are summarized as follows:
\begin{itemize}
    \item BaseSAFE is an emulation platform offering zero-overhead Rust bindings for emulation, sanitizing, and fuzzing,
    \item The developed fuzzing toolkit can be used for any other baseband or kernel,
    \item We show the viability of automatic test-case inference through coverage feedback,
    \item We provide insights into a major baseband, MediaTek, and discuss bugs found with BaseSAFE,
    \item We fuzz layer 3 signaling message handlers, as some layer 3 messages are unencrypted and allow pre-authentication attacks,
    \item We provide a proof-of-concept use-case of BaseSAFE fuzzing RRC 
signaling messages and Non-Access Stratum (NAS) EMM messages in MediaTek 
basebands.
\end{itemize}
With BaseSAFE, we provide the groundwork for automated fuzz tests of low-level parsers in closed-source, embedded targets.

\subsection*{Availability}
BaseSAFE is built on open-source software.
Its source code and test cases are open-sourced at \url{https://github.com/fgsect/BaseSAFE}.

%% file: background.tex
\section{Background}

This paper can be seen as the intersection of two fields: cellular baseband research and fuzzing.
Because of this, we will give a thorough introduction to both topics, with the goal that the reader will be able to follow the paper, independent of the background.

\subsection{Fuzzing}

Fuzzing is a powerful way to detect vulnerabilities, especially in low-level code.
In recent years, fuzzing of desktop software and parsers using tools like~\emph{AFL} and its fork,~\emph{\AFL{}}, has become one of the main tools for automated analysis~\cite{aflplusplus}.
The fuzzer reruns the target with different input thousands of times per second, using heuristics to mutate the test cases.
A key factor to a fuzzers' success is feedback from the target, usually coverage feedback.
Thanks to coverage feedback the fuzzer knows if the last mutation triggered a, potentially vulnerable, path in the program~\cite{zalewski2016technical}.

Fuzzing of embedded systems is a challenging task, especially if feedback should be collected and memory corruptions should be detected quickly~\cite{muench2018you}.
A few years ago, fuzzing wireless stacks, firmware or a kernel, required complex setups.
They had to resemble real-world scenarios, like dedicated rogue access points~\cite{butti2008discovering} that are difficult to integrate into feedback-based fuzzing methodologies.
In contrast to user-land software, in bare-metal systems and firmware, any state change affects the whole system.
Recovering from crashes is oftentimes impossible.
%TODO: kAFL
\subsection{Unicorn}

\emph{Unicorn engine} (or just \emph{Unicorn}), a corner stone of BaseSAFE, is a fork of QEMU~\cite{unicornemu}.
Unicorn extends QEMU with an easy to use API, exposing functions like reading and writing memory, and hooking specific addresses and memory accesses with custom callbacks.
Unicorn offers bindings for a range of languages.
However to use Rust in BaseSAFE, we extended the 3rd party \emph{unicorn-rs} bindings, as no official Rust bindings exist~\cite{unicorn-rs}.

Unicorn supports a vast range of processor architectures, including MediaTek's baseband architectures, ARM and MIPS~\cite{unicornemu}.
This makes emulation of arbitrary code, even for embedded architectures, viable.

QEMU, as well as the forked Unicorn engine, work by performing the following steps for each new code location that needs to be run~\cite{qemutcg}:
\begin{enumerate}
    \item Check if the translation block (instructions to the next conditional jump) at this location were previously cached.
    \item If not cached, decode and lift the translation block at this address from the target platform's instruction set to \emph{Tiny Code Generator (TCG)}, the internal intermediate representation.
    \item Translate the \emph{TCG} to the host platform's instruction set.
    \item Cache the translated block.
    \item Store a mapping from source program counter to target program counter in an address lookup table.
    \item Execute the translated block.
    \item Repeat for the next discovered block.
\end{enumerate}

The translation blocks are similar to basic blocks by design~\cite{bellard2005qemu}.
Leveraging the correspondence between translation blocks and basic blocks, and because execution is handed back to the emulator after each run, it is possible to implement an instrumentation similar to the compile-time instrumentation, using the program counter as feedback on each new basic block.
\AFL{} offers this instrumentation with \textit{QEMU mode}.
It leverages a patched version of QEMU that reports executed branches back to AFL~\cite{qemumode}.
After a new basic block is translated, the fork server's parent is informed that the block has been translated to ensure every block is translated only once.
The control returns to the emulator after each block, which is extended with calls to \texttt{afl\_maybe\_log}.
This call fills a shared memory section, passing the instrumentation information to AFL.

We built up instrumentation for BaseSAFE very similar to the AFL-QEMU instrumentation.
Translation blocks are cached in the parent process of the SafeBASE forkserver to increase the throughput of future runs.
AFL merely has to start the harness and generate inputs.
The forked Unicorn used for BaseSAFE makes use of the same concepts, tightly.

\subsection{Cellular Baseband}
%TODO: why we chose baseband? - hard target, mostly manual analysis, touching
%every packets processed in the smartphone

Every modern smartphone has multiple types of processors. Apart from the 
application processor, which runs the mobile operating system (OS), modern smartphones use 
independent baseband processors. The baseband processor handles all cellular 
communication. Even though the application processor and baseband processor are 
physically integrated into a single processor die, they are logically separate.
Baseband processors run a different operating system, usually a real-time OS (RTOS), 
not a full-featured system like the main processor does. Depending on the make, 
the method for inter-communication between both processors varies. As 
smartphones require faster mobile internet speeds, the internal interconnect bus 
requires more bandwidth, so sometimes processors also utilize shared memory 
between application and baseband processor. Thus, in some cases, attacking a 
smartphone OS via its baseband is also possible~\cite{Miru2017}.

One of the main tasks of the cellular baseband is to identify the cellular 
networks, to authenticate, and to connect to the correct one. To achieve this, 
the baseband scans for the radio signal coming from the cells and decodes 
network information messages from \textit{System Information Block (SIB)} 
messages to identify the network. After identifying nearby cells, it will 
connect to the strongest cell and perform the registration procedure. Once 
everything is done, the cellular services are available. After a connection has 
been established, the baseband provides mobile telephony and data services to 
the mobile OS.

%TODO: attack model
As pointed out by Rupprecht et al.~\cite{Rupprecht2017}, various attacks are 
possible just with an attacker-controlled rogue base station. One of the 
earliest attacks targeting a baseband was presented by 
Weinmann~\cite{Weinmann2012}. Since there is no reliable way to identify whether 
the base station is genuine or not, an external attacker can set up a rogue base 
station (also known as IMSI catchers~\cite{Strobel2007}) and send signaling 
messages like the legitimate operator, without being noticed by a user. It is 
possible for an attacker to inject modified signaling messages by utilizing the 
modified open-source software (e.g. Osmocom~\cite{osmocom} for 2G, 
OpenBTS-UMTS~\cite{openbts-umts} for 3G, srsLTE~\cite{srslte} for 4G) or the 
software of a commercial base station. With the wider availability of 
software-defined radio (SDR) devices, the total cost for the rogue base station 
setup became feasible to the attacker nowadays. Broadcasted signaling messages 
are processed by devices without explicitly establishing the connection, while 
dedicated signaling messages are available only after establishing a connection 
to the base station.

%% file: related.tex
\section{Related Work}

To the best of our knowledge, no fuzzing API for basebands exists so far.
This section discusses related work in fuzzing and cellular security.

\subsection{Emulator-Based Fuzzing}

Different ways to use emulation for snapshot and kernel fuzzing exist.
Notable examples include \emph{TriforceAFL}~\cite{triforce} by Hertz and Newsham, as well as \emph{kAFL} by Schumilo et al., both extending AFL's QEMU mode to fuzz whole virtual machines.
In both cases, the fuzz driver communicates with QEMU through additional hypercalls~\cite{kAFL}.
The current state of the art kernel fuzzer for desktops, \emph{Syzkaller}, uses VMs as well. A user-land stub inside the VM triggers the kernel via syscalls~\cite{syzkaller}.
Likewise, \emph{Unicorefuzz}, by Maier et al., fuzzes kernel functions in a QEMU-based emulator, Unicorn engine.
In contrast to the other kernel fuzzers, Unicorefuzz only maps memory actively used in the fuzzed function and runs this single snapshot continuously~\cite{ucf}, forwarding each newly accessed memory from the targets using \emph{avatar\textsuperscript{2}}~\cite{Avatar2}.
Schumilo et al. take it one step further, fuzzing hypervisors instead of kernels, in a similar fashion inside a custom hypervisor~\cite{hypercube}.

\subsection{Baseband Research}

% \todo{Cellular Fuzzing Papers (anno 2010)}
% \todo{maybe this?}
% \todo{\url{https://ieeexplore.ieee.org/abstract/document/6823894}}

% \todo{Name Sect 2FA SMS bugs through fuzzing?}
% \todo{\url{https://static.usenix.org/events/sec11/tech/full_papers/Mulliner.pdf}}
% \todo{\url{https://www.blackhat.com/presentations/bh-usa-09/MILLER/BHUSA09-Miller-FuzzingPhone-PAPER.pdf}}
% \todo{\url{https://link.springer.com/chapter/10.1007/978-3-642-39235-1_9}}

The root cause of previously proposed attacks targeting a cellular baseband 
can be traced back to the specification and the implementation. The errors in 
the standard allowed various forms of privacy leakage such as cellular 
subscriber tracking~\cite{Shaik2015} and data 
eavesdropping~\cite{Rupprecht2019}. While there had been previous works on 
individual baseband bugs~\cite{Mulliner2011, Rupprecht2016, Park2016}, we are 
not aware of a systematized approach on identifying the baseband bugs based on 
fuzzing.

Features of a baseband that were proven exploitable in the past are SMS messages and AT commands.
The 3GPP SMS specification~\cite{3gpp.23.040} defines more than just a text SMS, and 
processing of the SMS had been exploited by previous researchers.
Mulliner and Miller~\cite{Mulliner2009} presented the bugs related to the SMS processing of mobile basebands, and Mulliner et al.~\cite{Mulliner2011} tested them 
over-the-air for multiple types of devices.
These bugs are also affecting SMS-based applications, such as one-time password~\cite{Mulliner2013} and SIM toolkit~\cite{Alecu2013}.
While mobile messenger services are replacing SMS, they are part of the cellular standard and SMS functionality exists in smartphones.

Originally designed for controlling dial-up modems, AT commands are also used in some modern smartphone basebands~\cite{3gpp.27.007} as a part of inter-processor communication.
As a result, by sending a malicious AT command it is possible to exfilterate information from it or crash it unknowingly from the mobile operating system.
Example of AT command handler fuzzing includes the work by Tian et al.~\cite{Tian2018} and ATFuzzer by Karim et al.~\cite{Karim2019}.

Other work on fuzzing the components of smartphone includes PeriScope by Song et al.~\cite{periscope} targeting device drivers and Hay~\cite{Hay2017} targeting Android bootloaders.

The analysis of a baseband firmware is not well-researched: Qualcomm, one of 
the major baseband manufacturer, uses a custom in-house architecture named 
Hexagon. Although Qualcomm provides SDKs for the Hexagon processor, few 
well-known disassemblers integrate it, one notable example being 
GSMK's IDA Pro plugin~\cite{gsmkhexagon}. This further limits the research 
using other tools. Nevertheless, most basebands use standard architecture, at 
least for the main, (non-DSP) processor. Golde et al.~\cite{Golde2016} and 
Miru~\cite{Miru2017} present a baseband firmware disassembly using industry-standard tools, for basebands using off-the-shelf architectures such as ARM and 
MIPS.

There had been attempts on fuzz testing cellular protocol implementations. 
Johansson et al.~\cite{Johansson2014} proposed a cellular protocol fuzzing 
framework, which is integrated into the existing telecommunication testing 
infrastructure. Similar commercial services exists, such as P1 Telecom 
Fuzzer~\cite{p1tf}. Hussain et al. proposed 
\emph{LTEInspector}~\cite{Hussain2018} and \emph{5GReasoner}~\cite{Hussain2019}, 
which utilize formal analysis on the 3GPP specifications to test the 
implementation of basebands. Their approach is based on a formal analysis using 
the cellular specification, translated into a machine-readable form. 
\emph{LTEFuzz} by Kim et al.~\cite{Kim2019} uses predefined test cases to 
identify implementation problems of a baseband. \emph{SpikerXG} by Hernandez et 
al.~\cite{Hernandez2019} fuzz firmware of Android devices and propose an 
analysis platform for further rehosting and analysis will be possible, taking a 
big step towards automated baseband analysis. Prior publicly known baseband 
fuzzing setups fuzzed leaked binaries compiled for host system~\cite{Miru2017}. 
This is a valid approach but only feasible if object files are available, not 
for off-the-shelf firmware blobs.

%% file: rehosting.tex
\section{Rehosting and Fuzzing}

\begin{figure}
    \Description[BaseSAFE overview]{This depicts how BaseSAFE uses afl++ to fuzz-test firmware. Each snapshot is emulated in Unicorn engine. Feedback is reported back to AFL.}
    \centering
    \includegraphics[width=\linewidth]{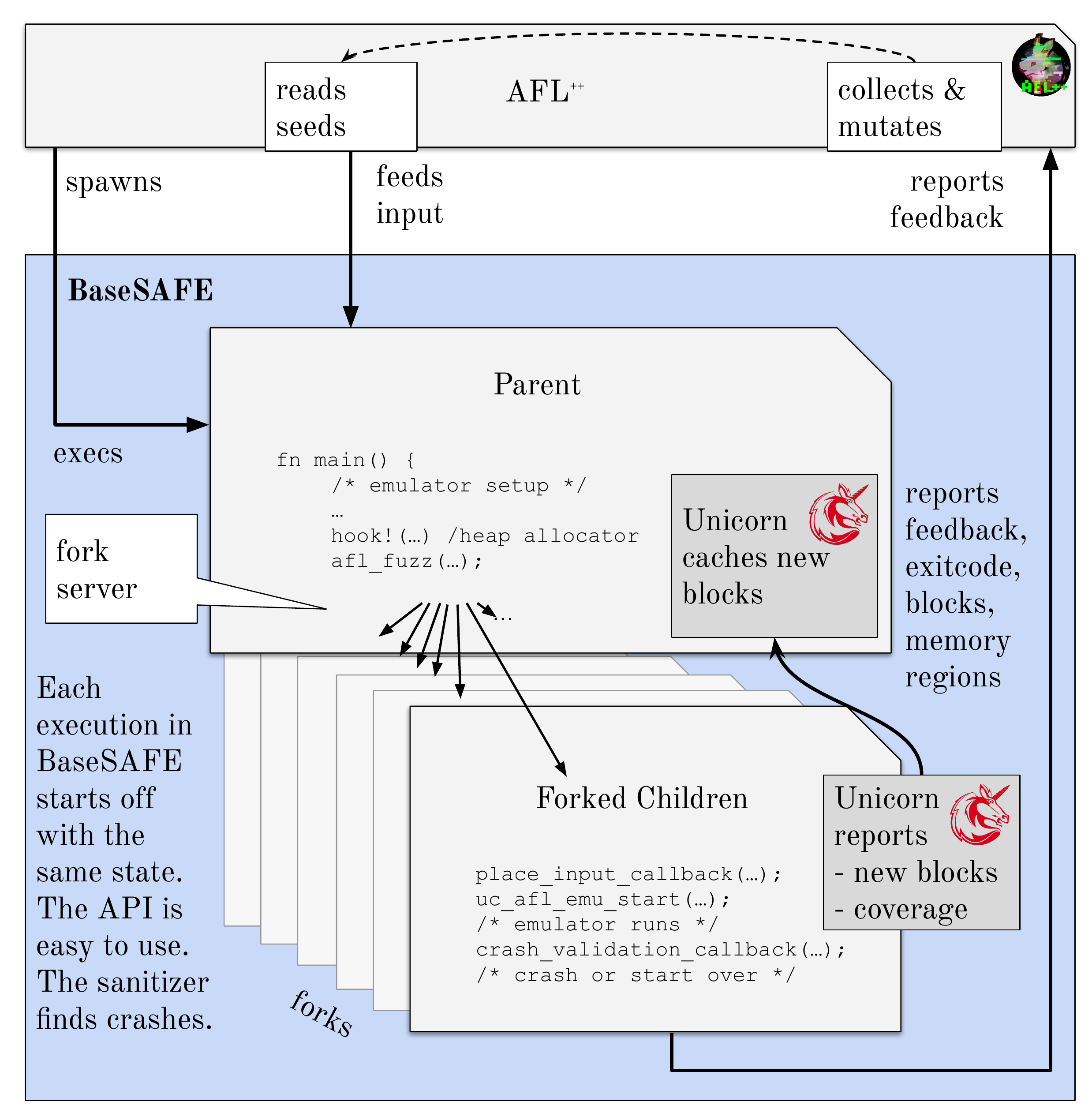}
    \caption{BaseSAFE}
    \label{fig:basesafe}
\end{figure}

BaseSAFE is a platform to build baseband fuzzers on.
The novel core components of BaseSAFE are an API to hook up the emulation toolkit with the maintained AFL fork \AFL{}~\cite{aflplusplus} in a very quick and flexible manner, as well as various useful components like a heap sanitizer.
To enable high-speed fuzzing, BaseSAFE uses emulation, built upon \emph{Unicorn 
engine}, a popular CPU emulator~\cite{unicornemu}.
This means, fuzzing of the baseband firmware does not take place on the phone 
hardware, but instead interesting functions are fuzzed directly inside Unicorn 
engine. The Rust API makes it easy to write powerful high-performance hooks, 
abstracting away potential hardware interactions, interrupts, etc. This allows 
emulation and fuzzing of the important parts of a firmware blob on fast desktop machines. Of course, 
different targets do still need manual setup, coding, and reverse engineering. In 
the following, we will discuss the usage and benefits of BaseSAFE.

\subsection{Fuzz API}

In this section, we discuss the BaseSAFE API.
Unicorn is extended with an AFL-specific API to enable easy fuzzing, and an extra API to hook the operating system's heap sanitizer.
A high-level overview of the BaseSAFE procedure is depicted in Fig.~\ref{fig:basesafe}.
The API of BaseSAFE goes beyond previous emulators, such as AFL Unicorn by Voss~\cite{aflunicorn} which did not offer interactions with AFL. Interactions with AFL are required to kick off the fast persistent mode, but simply always started fuzzing after the first instruction.
As execution left the emulator after the first instruction to read AFL input, all Unicorn translation caches were flushed constantly.
On top, it only worked for harnesses written in the interpreted and garbage collected language Python.

The Unicorn engine API includes functions to set page mappings, read and write memory and registers, add hooks, as well as start and stop execution with different conditions.
BaseSAFE makes them available via Rust.

We extended this part of BaseSAFE with the following methods, tailored for fuzzing:

\subsubsection*{\texttt{afl\_forkserver\_start}}

After the initial setup of the test case is done, the fuzz harness can call \texttt{afl\_forkserver\_start}.
This kicks off the forkserver logic.
The baseband is kept in the same state in the parent process, each fuzz test case is executed against a forked copy of the emulator.
So a call to \texttt{afl\_forkserver\_start} effectively freezes the current state of prior to fuzzing run and at the same time tells the attached \texttt{afl-fuzz} process to generate inputs.
After the forkserver started, the harness should read the input for this run from \AFL{} and place it into the appropriate location in the target's memory.
As the fork is copy on write, the test case data will be reset at the end of the run.
Once this happens, the parent process requests the next test case.
Furthermore, the forkserver contains a caching mechanism for Unicorn's JIT, inspired by the AFL QEMU mode:
for each uncached basic block the child encounters, the child will
\begin{enumerate}

    \item Decode the basic block from the target architecture (for MTK, this is either 32 bit ARM or MIPS).
    \item Add instrumentation to the block, namely register each jump from one location to the next in a shared map that will be evaluated by \AFL{} to generate further inputs.
    \item Notify the parent process about the current block address and flags.
    \item Run the basic block.
    \item Continue with the next block. If it is already cached, patch in a direct jump.

\end{enumerate}
Once notified about a new block address by the child, the parent process will also translate this block.
As the parent mirrors the child's translation, this block will already be present in the block cache for the next test case.
This concept was carried over from AFL's QEMU mode, albeit with speed improvements: whereas QEMU mode caches inside of the emulator, BaseSAFE patches it into the translated block itself, reducing the need for indirect jumps, a method first proposed by Biondo~\cite{improvingaflqemu}.

\subsubsection*{\texttt{afl\_next}}

As the \texttt{fork} syscall is heavyweight and, therefore, rather slow, BaseSAFE offers a faster alternative: persistent mode.
For targets like single parsers, for example \texttt{errc\_event\_handler\_main}, afl's persistent mode can greatly improve fuzzing speeds.
Instead of exiting and reforking, the child process resets its state internally, resets the needed stack, memory, and registers, and then calls \texttt{afl\_next} to inform the BaseSAFE parent and afl-fuzz about the end of a single fuzz run.
\AFL{} will then place the next test case and the child can call the fuzzed target again.

\subsubsection*{\texttt{afl\_emu\_start}}

In contrast to the \texttt{emu\_start} function offered by the core Unicorn emulator, \texttt{afl\_emu\_start} of BaseSAFE takes multiple exit addresses.
This is required if the target may not always return at the end of a function, for example if error conditions trigger.
On top, it will not clear the translation block cache, containing the JITted basic blocks after execution, as it is the case with the emulator function in Unicorn.
This allows us to reuse the cache for consecutive forks, as discussed in \texttt{afl\_forkserver\_start}, as well as persistent mode, discussed in \texttt{afl\_next}.
Recompilation of the basic blocks is therefore not needed.
Fuzzing speeds are high once the emulator has encountered and translated most of the blocks.

\subsubsection*{\texttt{afl\_fuzz}}\label{sec:aflfuzz}

Instead of manually specifying the fuzzing logic, such as reading AFL's input, catching Unicorn exceptions, or looping back to the beginning for persistent mode, the developer may choose to use the all-in-one function \texttt{afl\_fuzz}.
After setting up the baseband inside the emulator, the fuzz function takes over all necessary steps to fuzz, including all of the functions mentioned above.
Its signature can be seen in Listing~\ref{lst:aflfuzz}.

The function \texttt{afl\_fuzz}
\begin{enumerate}

    \item Loads the current test case input from AFL.
    \item Calls the \texttt{place\_input\_callback}, in which the harness should write the input into the emulator memory at the appropriate position. 
    For persistent mode, the emulator has to reset additional state changes in this step.
    \item Runs the emulator until execution reaches one of the exits, a hook crashes execution or an illegal state occurs in the emulator.
    \item Checks the Unicorn emulator error conditions and (optionally) calls the \texttt{crash\_validation\_callback}, allowing us to implement custom sanitization routines.
    \item Starts from the top if persistent mode is enabled and the counter did not expire.

\end{enumerate}

\begin{minipage}{\linewidth}
\begin{lstlisting}[captionpos=b,frame=single,language=Rust,label=lst:aflfuzz,caption=Function Signature of \texttt{afl\_fuzz} in Rust.]
pub fn afl_fuzz<F: 'static, G: 'static>(
    &mut self,
    input_file: &str,
    input_placement_callback: F,
    exits: &[u64],
    crash_validation_callback: G,
    always_validate: bool,
    persistent_iters: u32)
                -> Result<(), AflRet>
  where
    F: FnMut(UnicornHandle<D>,
             &[u8], i32) -> bool,
    G: FnMut(UnicornHandle<D>, uc_error,
             &[u8], i32) -> bool           {
\end{lstlisting}
\end{minipage}

In the \texttt{input\_placement\_callback} the harness writes the input test case to the emulator memory.
The callbacks both get pointers to the input of the current test case, provided by AFL, as well as the persistent iteration index if it is required.
Furthermore, the \path{crash_validation_callback} will get the exit code from Unicorn, which will indicate errors caught during emulation, such as out-of-bounds memory accesses.
Depending on its return code, the exit condition may be considered interesting, e.g., a crash or hitting sanitization, in which case the AFL process gets the information forwarded over an Inter-Process Communication (IPC) mechanism.
The fuzz function also takes a list of exits at which emulation will stop, a flag whether the validation callback should also be called without a Unicorn error condition and an additional u32 counter, indicating if---and how often---persistent mode should loop before forking again.

\subsubsection*{Debug Tracing}

While execution has to be as fast as possible for fuzzing to allow a large amount of test cases to be evaluated, requirements when triaging a crash are vastly different.
Instead of high speed, the user wants a good understanding of the execution path taken and encountered error cases, and hence needs as much context as possible.
For this, BaseSAFE features a debug mode that will output all disassembled instructions and register values during execution.
As long as debug outputs are enabled, our harness also prints all output directly generated by the baseband, such as logs generated by \texttt{dhl\_trace} and assert messages.

\subsection{The MediaTek Baseband}
% whttps://topic.alibabacloud.com/a/on-cottage-phone-and-android-quot7quot-mtk-mobile-phone-software-system_1_12_30805610.html

\begin{figure}
    %https://docs.google.com/drawings/d/1rEK1VzyROztmXhkgo7z7h3ZbZUrsITGz-iEtDpF_jpA/edit?usp=sharing
    \Description[Mediatek Overview]{This shows the overview of the message layers of the OS. We are mostly intrested in layer 3}
    \centering
    \includegraphics[width=\linewidth]{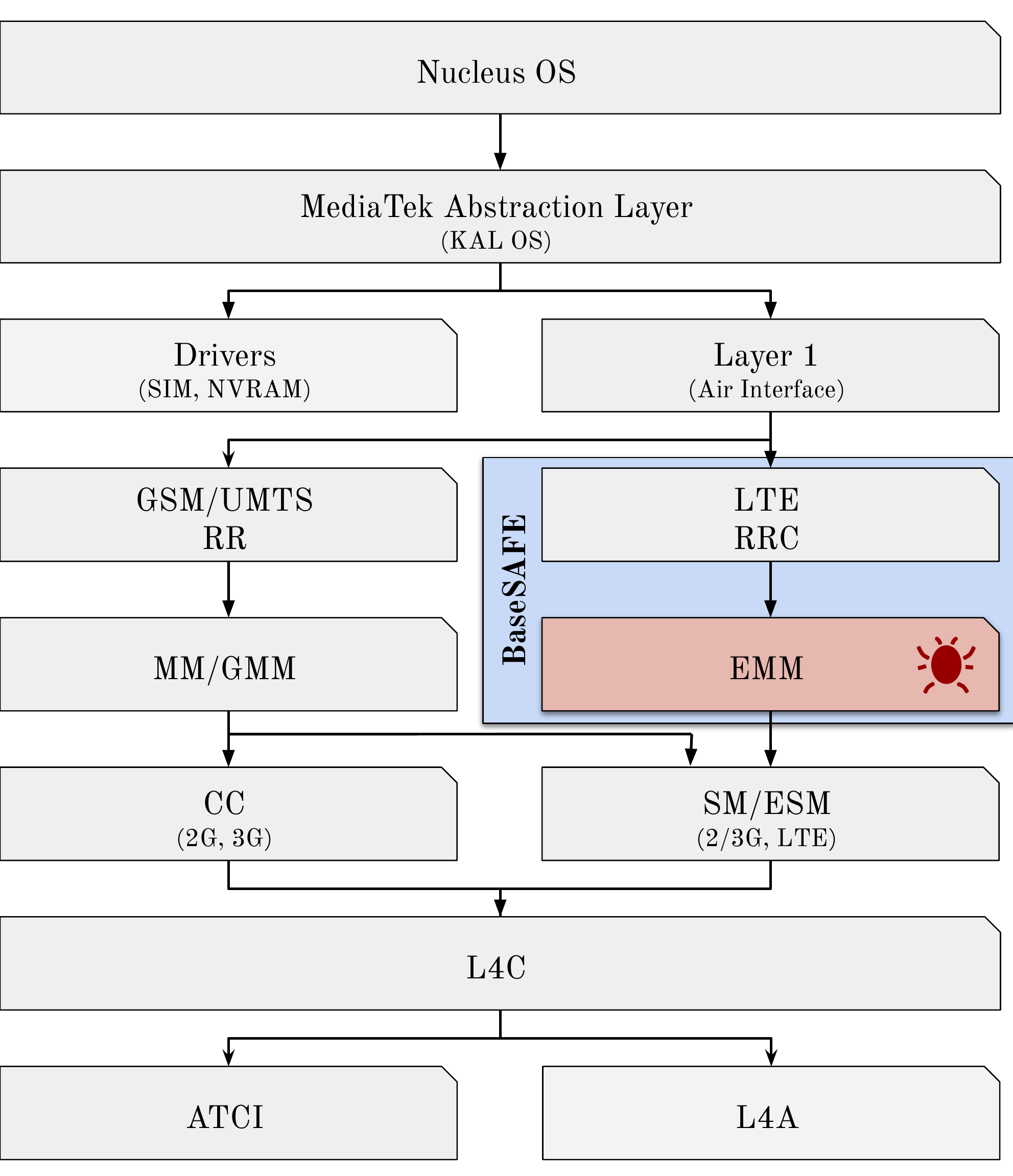}
    \caption{MediaTek Baseband Architecture.}
    \label{fig:mtk_overview}
\end{figure}

We chose a MediaTek baseband as a target to implement a proof-of-concept fuzzing harness, 
proving the usability and viability of BaseSAFE. First, we will present an 
overview over the reverse-engineered baseband.

For the course of this work, we focus on the \emph{HTC One E9+}, using 
MediaTek's Helio X10 (MT6795) processor launched in 2015~\cite{wikichipx10}. 
Aside from HTC One E9+ and M9+, this processor is also used in some smartphones 
around 2015, including Sony Xperia M5 and Xiaomi Redmi Note 2 and 3. We were 
able to access the unencrypted baseband firmware using the tool from 
Miru~\cite{Miru2017}, which is targeting the same processor but tested in a 
different device. MediaTek's baseband firmware consists of two parts: ARM and 
DSP. The DSP firmware controls the lower layers, including modulation and 
demodulation of the over-the-air signal. The ARM firmware controls the upper 
layers, including the processing of the signaling messages and interconnection with 
the application processor. We are focusing on the ARM part of the baseband 
firmware, as this is the place where cellular control plane messages are 
processed and, therefore, can be controlled by an external attacker.

An overview of the internal structure of the ARM firmware is depicted in 
Fig.~\ref{fig:mtk_overview}.
Most of this MediaTek \emph{MTK} baseband OS was likely carried over from their prior feature phone OS, providing the same task model and IPC mechanisms~\cite{freetibet}.
The OS is generally following the layer model of the 
cellular network. The lowest layer is a Nucleus RTOS kernel with a MediaTek 
abstraction layer above it. Drivers are interacting with external entities used 
by a baseband, such as NVRAM, used to store baseband configurations, and SIM 
cards. Layer 1 and 2 are not clearly separated within the MediaTek baseband 
firmware. Layer 3 consists of multiple protocols related to the separate 
functionalities of the cellular network: radio resource control (RR, RRC), 
mobility management (MM, GMM, EMM), call control (CC), and session management 
(SM, ESM). Above this lies an application layer, layer 4, whose implementation 
is MediaTek-specific. It consists of a command interpreter (ATCI), a control 
entity (L4C) and an adaption layer (L4A) and interacts with the mobile OS 
running on the device's application processor.

We are focusing on LTE RRC and EMM messages, as part of these messages are 
normally used for identifying a base station and exchanged upon establishing a 
connection between smartphone and network. As such, vulnerabilities in handling 
these messages enable a pre-authentication attack.

Imagination Technologies--then owner of MIPS--announced that MediaTek is adopting MIPS as the architecture for their baseband~\cite{mipsmtk}.
This is a departure from the ARM-based core that we analyzed as a proof-of-concept for this work.
By analyzing the smartphone's baseband firmware images by chipset using binwalk opcode analysis and Ghidra, we also found that MediaTek used ARM in their basebands released before 2017 (Helio P25, X10, X27) and switched to MIPS around mid-2017 (Helio P23, P70, P90, X30).
Our reverse engineering indicates that the parsers we are fuzzing are close to identical in MIPS-based MediaTek basebands.
This is to be expected, as code reuse across architectures minimizes the development costs.
Of course, the proof-of-concept harness of BaseSAFE can easily be ported to MIPS, as Unicorn itself supports the architecture, although it might not provide novel insights due to the aforementioned code reuse.
Likewise, the MTK toolkits are rather standardized for both ARM and MIPS.

%\todo{MTKLogger - but not sure whether we really need this...}\\

\subsection{Nucleus and MTK Firmware IPC}\label{sec:nucleus}
\begin{figure}[ht!]
\centering
\Description[Input placement]{The input is placed at specific points of the ILM struct}
\includegraphics[width=\linewidth]{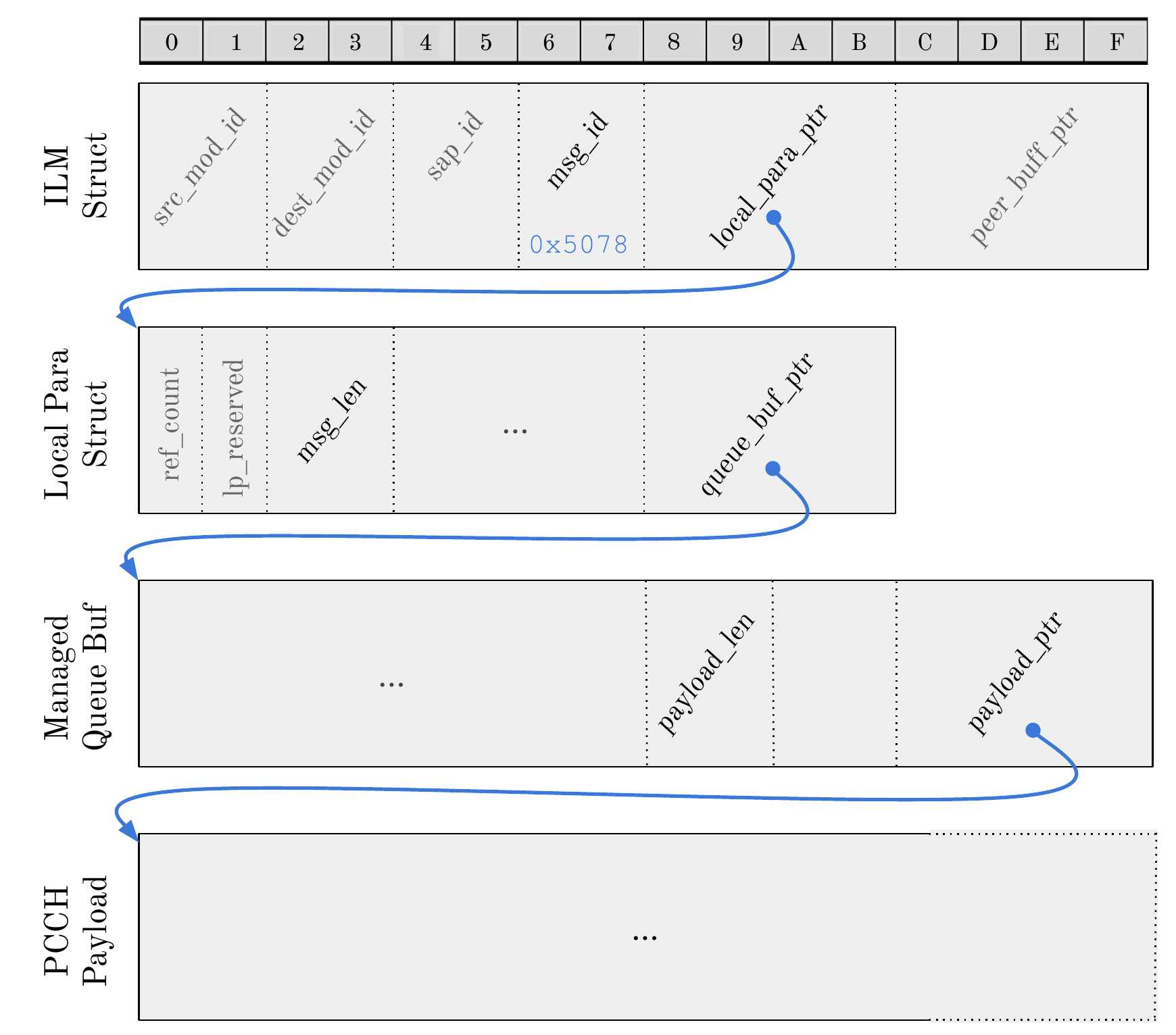}
\caption{Structure of MTK ILM, encapsulating a PCCH Message including pointers needed for correct input placement.}
\label{fig:structs}
\end{figure}

The ARM firmware we used as an example application for BaseSAFE is based on the Nucleus RTOS kernel, the libraries indicate a kernel version of 2.x.
The Nucleus RTOS is a non-preemptive realtime OS with a queue-based IPC mechanism.
The main method for the different parts of the MTK firmware to communicate with each other and forward packages to higher layers are these queues.
Modules can use the \texttt{do\_send\_msg} function or its wrappers to send so-called Inter Layer Messages (ILMs) to the internal or external queue. The function proceeds to call \texttt{kal\_enque\_msg} in the Kernel Abstraction Layer, which uses Nucleus primitives to pass the MTK message representation to the Queue Management Unit (QMU) of the kernel.
On the other end, modules use \texttt{msg\_receive\_extq} or \texttt{msg\_receive\_intq}, wrapping \texttt{kal\_deque\_msg}, in order to receive ILMs for further processing.
Each queue item, i.e. ILM, contains an identifying message ID and destination module ID.
The MTK firmware uses this information tuple to route the queue entry internally and start the correct handler.
Message IDs then provide hints on how to process messages, e.g., parsers handling incoming (physically external) messages could be informed on which decoder to use.
Additionally, ILMs handle domain-specific manifestations of local parameters and peer buffers, carrying various forms of information.
When an ILM is initially constructed, the functions \texttt{construct\_int\_peer\_buff} and \texttt{construct\_int\_local\_para} can be used to allocate and populate the respective buffer. 
As a next step, the \texttt{get\_int\_ctrl\_buffer} wrapper calls into \texttt{kal\_get\_buffer} which communicates with the Nucleus Partition Manager to return a fresh memory chunk. 
Both buffers have reference counts, a module currently using one of them would signal its demand by calling e.g. \texttt{hold\_local\_para} and thus increasing the counter.
Specific \texttt{free} wrapper functions, namely \texttt{free\_int\_local\_para} and \texttt{free\_int\_peer\_buff}, first check the reference count and free the corresponding buffer only if it is 0, otherwise it is decremented by 1.
The memory layouts of the ILM and local parameter structures are depicted in the upper part of Fig.~\ref{fig:structs}.
The MTK firmware also implements an own queue management unit for the buffer management of incoming messages and wraps received messages in metadata, as depicted in the diagram in the lower part of Fig.~\ref{fig:structs}.

\subsection{Selective Emulation}

Instead of emulating the whole baseband, we selectively emulate single parsers. 
The selective emulation is single-threaded and only spawns one process. This 
allows us to emulate quickly and with zero false-positives. However, this way we 
have to be more selective about the portion of the baseband we fuzz---as 
manual effort is required for each harness.

As illustrated in Fig.~\ref{fig:mtk_overview}, we are focusing on layer 3 
control plane signaling messages, in particular, 3G~\cite{3gpp.25.331} and 
LTE~\cite{3gpp.36.331} RRC and NAS EMM~\cite{3gpp.24.301, 3gpp.24.008} 
messages. They are relatively easy to modify with a fake base station compared 
to lower layers. Some of these messages, namely SIBs and \emph{Paging} messages 
are not encrypted and decoded automatically upon knowing the cell even without 
connecting to them. Even though these messages have been a target for various 
previous works, they still remain valid as a fuzzing target for various reasons. 
Signaling messages are encoded in a binary-based format such as ASN.1 and CSN.1, 
and correctly implementing a parser for those protocols can be bug-ridden, while 
modern ASN parsers are oftentimes autogenerated, making these parts less of a 
pressing issue. Even though there are compliance tests for the standards, these 
are developed towards the interoperability among multiple vendors, not towards 
the implementation problems of an individual baseband. The large amount of more 
than one thousand different signaling messages for LTE RRC alone we were able to 
deduce through fuzzing is likely not completely covered by compliance tests. 
Hence, it is likely that individual baseband manufacturers can make individual 
mistakes, which may not have been filtered by their internal testing.

To this end, we identify the handling functions of the aforementioned 
signaling messages of the baseband firmware structure with minimal user input. 
We start from the downlink RRC messages of the cellular network, which is 
categorized as follows:

\begin{itemize}
\item PCCH: Delivers paging messages. Plaintext message without authentication.
\item BCCH DL-SCH: Delivers SIBs to identify the network. Plaintext message 
without authentication, baseband automatically receives the contents upon cell 
discovery.
\item CCCH: Initiates a connection between the phone and the base station. 
Plaintext message without authentication.
\item DCCH: Dedicated channel between the phone and the base station. Integrity 
protected and optionally (but usually) encrypted.
\end{itemize}

In addition to RRC messages, NAS EMM messages manage the registration of a 
phone and mobility. Some of the messages are exchanged without encryption, therefore 
malicious EMM messages could be used as pre-authentication attack.

We have collected signaling messages from the real network and phone using 
SCAT~\cite{scat} and use them as seed inputs to potential signaling message 
handlers.

The different decoders expect input buffers to be placed at specific offsets in multiple layers of internal structures.
For incoming signaling messages, new buffers are allocated by the MTK-internal QMU. These buffers contain administrative queue metadata as well as the size of the incoming message and a pointer to the payload buffer. 
A local parameter struct is populated with a pointer to the queue buffer and is being held by an Inter Layer Message struct. 
Finally, the ILM needs to contain the correct message ID, signaling the higher-level function into which ASN.1 decoder the incoming message should be passed.
Fig.~\ref{fig:structs} depicts the whole layout using a PCCH message as an example.
Correct placement of buffers, pointers and length fields can be handled in the \texttt{place\_input\_callback} discussed in Sect.~\ref{sec:aflfuzz}.

\subsection{Parser Deduction}

\begin{figure}[h!]
    \Description[Coverage Plot]{The Coverage is higher for the packages fitting the respective parser.}
    \begin{center}
        \includegraphics[width=\linewidth]{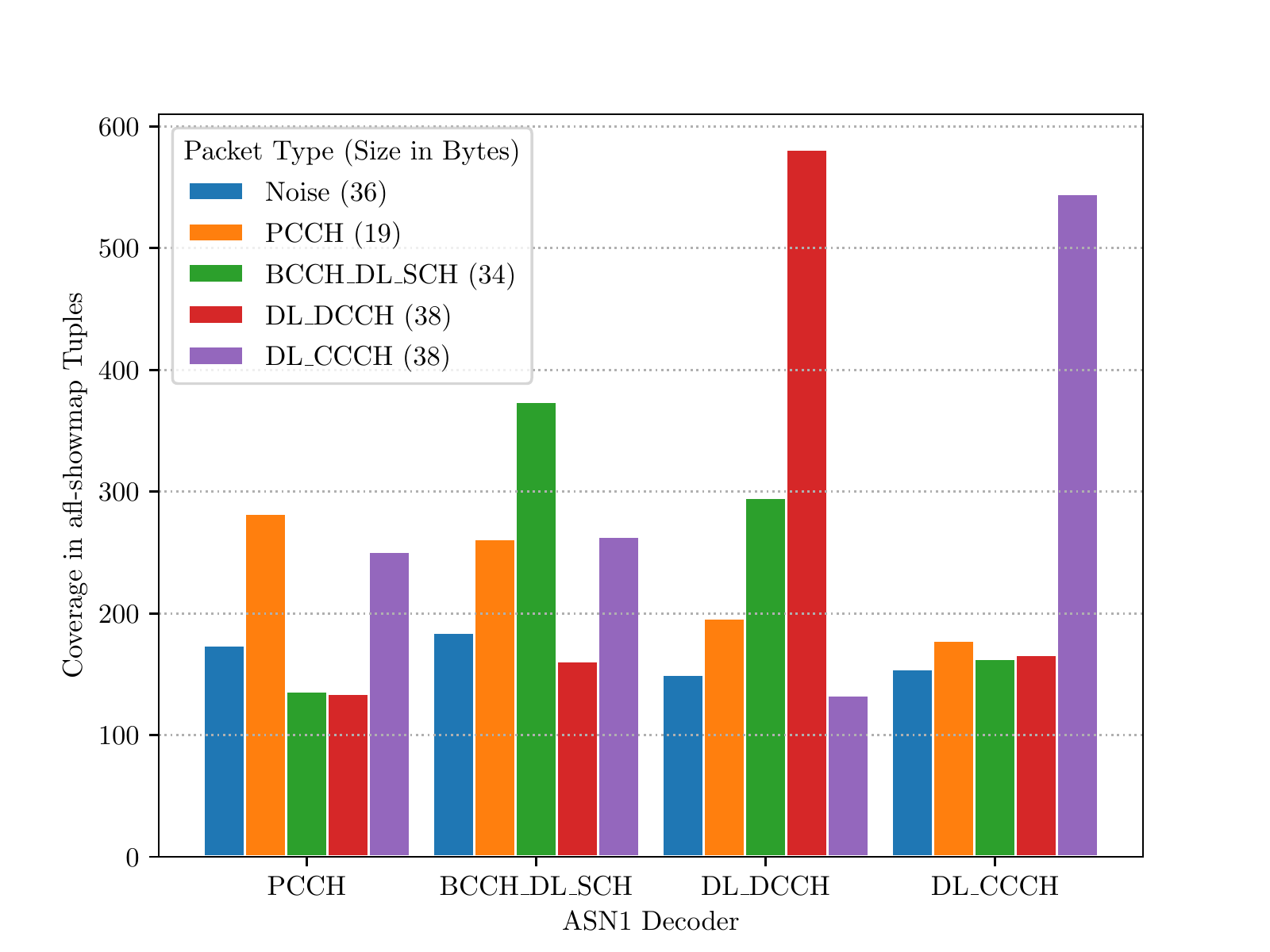}
    \end{center}
    \caption{Reported edge coverage of packages for each parser.
    The correct packages reaches the highest coverage in each.}
    \label{fig:coverage}
\end{figure}

% todo: check coverage on bigger PCCH packet (from afl generated stuff)
BaseSAFE can guide reverse engineers to select good fuzzing targets with a concept that we call \emph{parser deduction}.
After we collect signaling messages we are interested in using SCAT, we deduct the correct parser for the signaling message by evaluating the code coverage of different input functions when presented with the signaling message, compared with code coverage of unrelated packages.
For this, we feed valid signaling messages into all decoders by emulating valid IPC messages and log the coverage.
As shown in Fig.~\ref{fig:coverage}, the correct signaling message always reaches the highest coverage in the correct parser, as expected.
The correlation for PCCH is smaller than that of the other signaling messages, likely due to the fact that the valid PCCH message we used was rather small, with only 19 bytes. 
To introduce a baseline, 30 noise signaling messages with 32 to 42 random bytes and an average length of 36 were generated.
These were fed into the discussed decoders and achieve notably less coverage than the valid signaling message for each decoder, the average coverage can be seen in Fig.~\ref{fig:coverage}.
This method of parser deduction allows us to select an interesting and correct target to selectively emulate and fuzz.

\subsection{Rehosting the Baseband with Rust}

%TODO -> Same as API earlier?
Fuzzing speed depends on various factors.
Instrumenting a binary using a debugger degrades performance.
Recent advancements focus on speed of path finding, but also on sheer execution speed of the instrumentation through lightweight hardware features~\cite{fasterAFLPT,kAFL}, by leveraging optimized runtimes like QEMU block chaining~\cite{improvingaflqemu} or vectorized virtualization~\cite{vectoremu}.
Faster instrumentation could be possible.

Current research has shown that AFL's QEMU mode's performance can be improved by re-enabling QEMU's block chaining, which merges code blocks if one ends with a direct jump. It is disabled because it interferes with AFL's instrumentation: Merged blocks don't jump back into the emulator after every single contained block, so it effectively disables tracing direct jumps. The author injects the instrumentation code into the translated code, and thus can safely enable block chaining. Combined with proper caching this yields a speedup of 3-4 times the mainline QEMU mode~\cite{improvingaflqemu}. This patch could be ported to AFL-Unicorn, and could significantly reduce the performance gap to compiler-assisted instrumentation.

\subsection{Drop-In Heap Sanitizer}\label{sec:sanitizer}

Muench et al. classify the results of memory corruptions found through fuzzing in different categories.
In their book, only \emph{observable crashes} and \emph{hangs} are easy to track.
Especially in a fuzzing scenario, where the same test-case restarts over and over, the categories they classify as \emph{late crashes} and \emph{malfunctioning} are impossible to spot.
Of course, the last class, \emph{no effect}, is impossible to track down without address sanitization~\cite{muench2018you}.
A memory corruption itself and the subsequent use of the corrupted memory may often be far apart.
At this point the fuzzer already stopped the execution of this run and started the next iteration, removing all traces of the bug.
Thus, even when a crashing input is given, detecting the underlying memory bug is impossible.
We solve this issue by implementing a drop-in allocator that makes use of the emulator features to provide sanitization.

\subsubsection{How to Drop-In}

Usually, a drop-in allocator would be put in place, either during compilation or linked dynamically when the program is loaded.
Both options are not applicable for our use case: embedded firmware without source code, tool-chain, or knowledge about the linker.
While binary patching could be considered, similar to RetroWrite by Dinesh et al.~\cite{retrowrite}, adding functionality in the emulation layer is less fragile and works for all supported instruction sets.
The hooking functionality of the Unicorn engine inserts checks for conditions and callbacks directly into the JITted code.

In contrast to the QASan sanitizer for QEMU by Fioraldi, that patches each memory access in QEMU~\cite{qasan}, we add the instrumentation on top of the memory access hooks already offered by Unicorn.
With a hook in place, Unicorn engine emits checks for conditions like memory accesses and executed instructions.
If the check triggers, the placed hook is called from the JITted code directly, without stopping the emulation.
Using this feature, we overwrite the firmware's internal allocator, i.e. \texttt{kal\_get\_buffer} described in Sect.~\ref{sec:nucleus}, by hooking its address.
Whenever the hook triggers, we allocate memory in a previously mapped page and pass the location to the firmware by filling in the correct register.
After the hooked allocation, we increase the program counter to skip the firmware's actual allocator function call.

\subsubsection{Allocator Implementation}

The custom allocator offered by BaseSAFE itself makes heavy use of Unicorn hooks for sanitization.
Before emulation starts, a memory region large enough to handle all possible allocations during a single run gets allocated inside the emulated target.
Initially, an access hook is placed on the whole region.
Each time the hook triggers, a memory out-of-bounds access is detected.
The nature of the hook allows us to distinguish between reads and writes and allows us to log the current instruction pointer.

\paragraph{Allocation}
When the baseband's allocator would be called, the Unicorn hook calls our allocator instead.
The allocation chunk is allocated by removing our memory hook from a portion of the memory region of this size.
A canary region with varying size, depending on the size of the allocation, is left hooked between each allocated chunk.
The canary region is, again, protected by hooks.
This way the only heap corruptions we cannot spot are writes skipping the variable-sized canary regions and accessing another already allocated chunk.
Normally, all out-of-bounds accesses are detected.

\paragraph{Deallocation}
The deallocator hook places a new memory access hook on the previously allocated region.
Since the memory region will never be reused for this single run, all use-after-free bugs are detected by this hook.
In addition, the chunk size, being part of a bookkeeping structure, gets set to 0. Whenever \texttt{free} is being called on a chunk with a size of 0, a double-free is detected.
The managing structures are not part of the heap itself and are kept outside of the emulator and hence cannot be impacted by memory corruptions.
As the forkserver will reset the emulator memory for each emulation pass by reforking, we do not have to clean the custom heap manually, unless the persistent mode is in use, in which case the hooks are replaced completely.

%% file: evaluation.tex
\section{Evaluation}

In the following, we will first discuss results gathered with BaseSAFE, including memory corruptions.
Then, we discuss how BaseSAFE repackages PCAPs, and how we replay the test cases against a real MediaTek-based device over-the-air.

\subsection{Exhaustive Test Cases From LTE RRC}
We ran BaseSAFE on the LTE RRC test case for around one week with about 15k executions per second. Benchmarking the fuzz case on an Intel i7-6700 CPU @3.40GHz, the speed fluctuated around 1.5k executions per second on a single core.
Although no crashes could be found during the RRC tests, we were able to uncover slow paths.
However, this cannot be abused for denial-of-service because LTE RRC 
timers~\cite{3gpp.36.331} reset the internal parser if signaling messages 
were not received within the time limit.

% since the LTE RRC parser may kill parsing after a given timeout.\todo{REALLY? 
% Can we try this on the phone?}

\begin{figure}
  \Description[Wireshark Output]{Wireshark trace of one of the fuzzer-generated signaling messages triggering unique code flow in MTK}
  \centering
  \includegraphics[width=\linewidth]{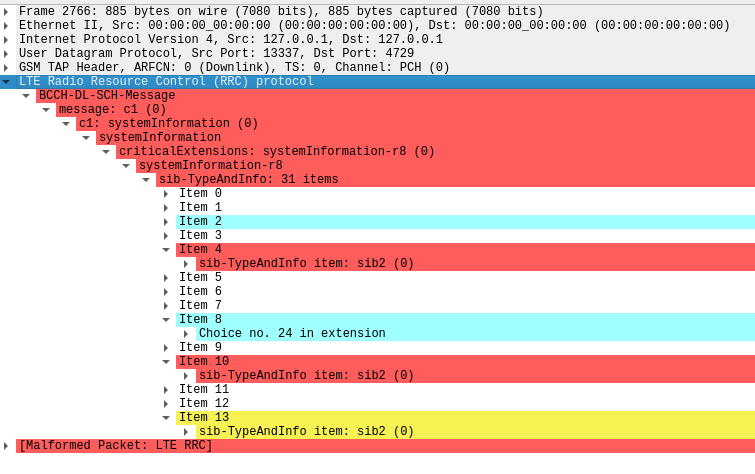}
  \caption{Wireshark trace of one of the fuzzer-generated signaling messages triggering unique code flow in MTK}
  \label{fig:wireshark}
\end{figure}

In total, after minimizing millions of test cases using \emph{afl-cmin} on the 
corpus and \emph{afl-tmin} on every single corpus-minimized input, 1388 unique test cases, leading to unique paths in the 
parser, were found, see Sect.~\ref{sec:redpill}. They will be released as part of BaseSAFE and can be reused 
to test the same message on other baseband firmware in the future. This number can still go up with longer fuzzing times. As these 
signaling messages were effectively outgeneraled by the parser, all of the 
messages are relevant for parsing, even if some behavior may not be specification-compliant, cf. Fig.~\ref{fig:wireshark}.

\subsection{Memory Corruptions in NAS EMM}\label{sec:emm}

Higher up in the LTE stack, in the NAS EMM parser, BaseSAFE was able to uncover out-of-bounds reads and writes. The position of the bug within the MTK architecture is shown in Fig.~\ref{fig:mtk_overview}.
After fuzzing the parser, the heap sanitizer, 
discussed in Sect.~\ref{sec:sanitizer}, successfully reported multiple such crashes with the same root cause.
The bug, first discovered by Grassi and Chen~\cite{grassi}, lies in \texttt{decodeEmergencyNumberList}, a function 
being called during the handling of \emph{Tracking Area Update Accept} and 
\emph{Attach Accept} messages, specified in the \emph{3GPP TS 
24.008}~\cite{3gpp.24.008} standard.
Here, an attacker-controlled length-byte is 
not checked against the correct message size, leading to an out-of-bounds 
\emph{read} from the received message buffer while copying it to the internal 
data structure. As the offset into the target buffer gradually increases by 
factor $0x2b$, an attacker would in addition be able to overflow it during the 
copying process and achieve an out-of-bounds \emph{write} on the heap.
\begin{minipage}{\linewidth}
\begin{lstlisting}[captionpos=b,frame=single,language=C,basicstyle=\ttfamily\tiny,label=lst:decodeEmergencyNumberList,caption=Loop in decodeEmergencyNumberList]
i = 0;
do {
  index = *ecc_number_list_struct;
  curr_item = ecc_number_list_struct + (uint)index * 0x2b;
  curr_item[2] = msg[i] - 1;
  data_start = i + 2;
  curr_item[1] = msg[i + 1 & 0xffff];
  j = 0;
  while (i = data_start & 0xffff, j < curr_item[2]) {
    data_start = i + 1;
    curr_item[j + 3] = msg[i];
    j = j + 1 & 0xff;
  }
  j = (uint)index + 1 & 0xff;
  *ecc_number_list_struct = (byte)j;
} while (i < length);
dhl_trace(TRACE_GROUP_1,0,DAT_001b8020,PTR_DAT_001b8024,j,length,msg_ptr);
\end{lstlisting}%
\end{minipage}
\subsection{The Red Pill}\label{sec:redpill}

To finalize the emulation evaluation, we need proof that the messages produced in the emulator are indeed valid messages for real basebands.
For this, we used two main methods.

\subsubsection{AFL Inputs to PCAPs}

After running for five days, the feedback-driven mutations of AFL generated a large corpus of potential inputs, with over 250k queued items.
Of course, many of these are rather similar.
As BaseSAFE works together with all \AFL{} tools, we can use a combination of AFL's minimization tools to reduce the amount of test-cases to 1388+ unique tests.
To arrive at this number:
\begin{enumerate}
  \item We ran \texttt{afl-cmin}, which loads the coverage map for each test case, then removes all test cases that only touch the same code paths from the list, keeping the smallest for each.
  This removes all files that do not reach new code paths, making sure each test case is actually relevant for the parser.
  \item We ran \texttt{afl-tmin} multi-processed on all remaining files. The test case minimization overwrites random chunks of the input file.
  If the coverage map stays the same, parts of the chunk are removed to check if the map also stays unchanged without these bytes.
  Similar heuristics are repeated until the file size is as minimal as possible---while maintaining the coverage.
\end{enumerate}

After the minimization process, we are left with a minimal set of inputs that still cover all possible branches of the original baseband parser.
In contrast to official test-cases, they may not be valid packages---but they will still trigger new conditions in the parser.
See, for example, the dissected packet containing signaling message in Fig.~\ref{fig:wireshark}.
To arrive at this dissection, and verify our method, we wrap the minimized test cases in a valid PCAP.
For this, BaseSAFE ships with a custom tool to wrap the test cases into a PCAP file.
The tool writes PCAP headers and then wraps the bytes each minimized test case into a GSMTAP packet.
The wrapped GSMTAP packets are decodable as a signaling message in Wireshark.
One of the generated test case decoded in Wireshark is presented in Fig.~\ref{fig:wireshark}.

\subsubsection{Replay Against Real Phones}

\begin{figure}
  \Description[Basestation]{A Base station setup, showing a laptop, a phone, antenna and the software defined radio USRP.}
  \centering
  \includegraphics[width=0.75\linewidth]{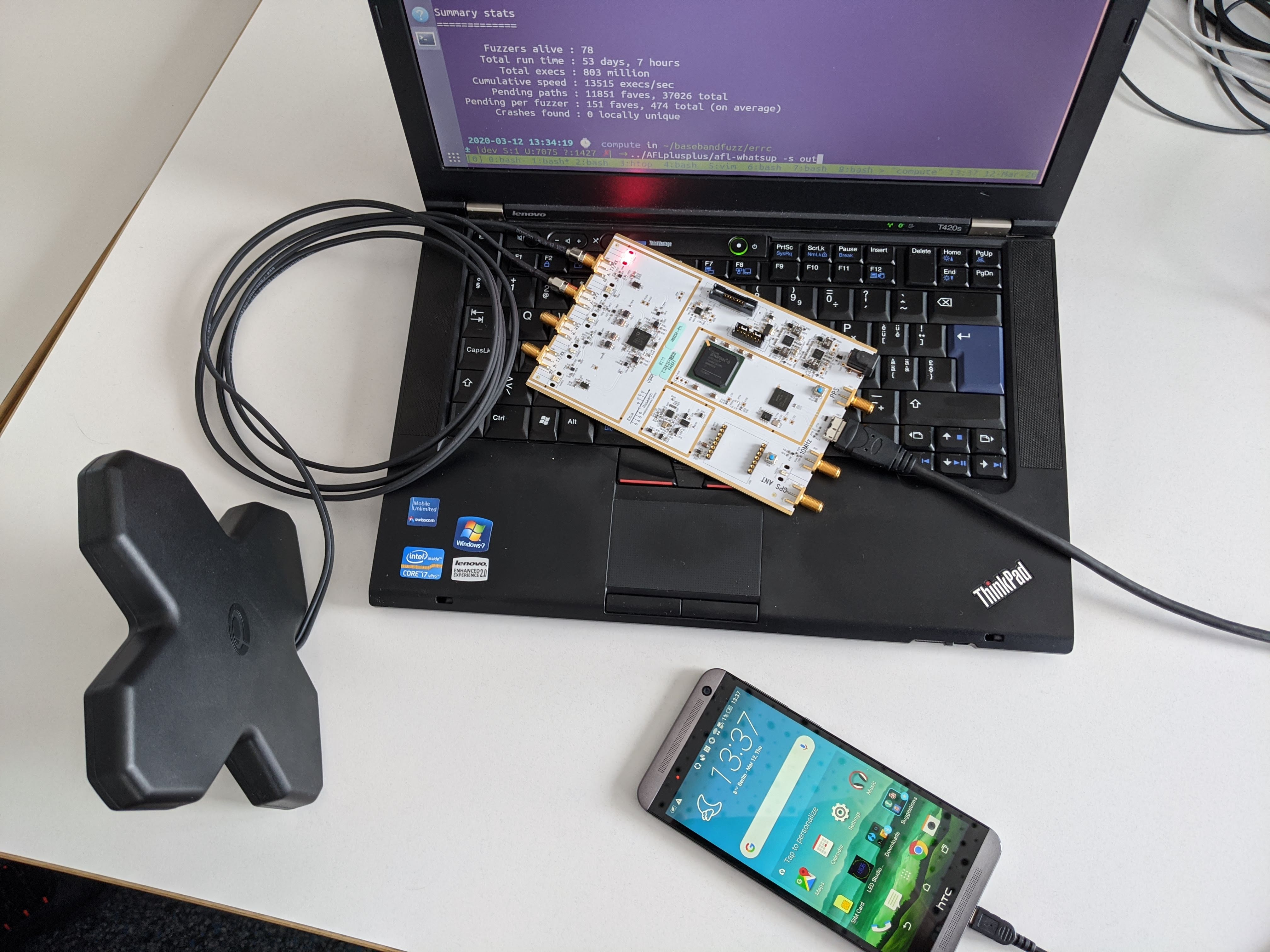}
  \caption{Base station setup with our test phone HTC One E9.}
  \label{fig:basestation}
\end{figure}

In order to try out responses in the real world and vet the behavior of BaseSAFE, we built a setup able to replay the signaling messages.
For this, we put up a rogue base station capable of injecting the messages at the correct time during the connection establishment procedure. 
We are using a software-defined radio and freely available application OpenLTE~\cite{openlte}, illustrated in Fig.~\ref{fig:basestation}. 
Along with this, we were able to analyze the baseband log output with the \emph{MTKLogger} system application.
The application is available on most MediaTek-based phones, although it is hidden from the user interface and needs to be accessed using special methods.

%% file: conclusion.tex
\section{Future Work}

BaseSAFE is an important first step towards fully automated vulnerability discovery on cellular basebands.
However, it only manages to cover a small portion, leaving many areas to explore further.

\subsection{Fuzzing for Logic Bugs}

Currently, our proof-of-concept harness for BaseSAFE mostly detects memory corruptions during parsing, except for the places where asserts were explicitly inserted by MediaTek into the baseband.
It will not discover any other bugs.
Through the insertion of additional hooks and more elaborately modeled execution flows, a variety of other bugs could be in the future.
For example, finding traces that disable a timer could lead to sustainable DoS and easier exploitation of bugs~\cite{Golde2016}.
On top, the parsed message structs, resulting from our LTE RRC fuzz test, could be handed to the consuming functions behind the baseband.
This could yield further bugs, as packages might pass the parser without corruptions, but the consumer could, in turn, blindly trust that input, just as we saw in Sect.~\ref{sec:emm}.
Using BaseSAFE hooks, a lot of other logic bugs could also be modeled.
For the baseband authentication, paths that reach an authenticated state without passing the necessary authentication functions can be hunted down this way.

\subsection{Collision-Free Coverage Tracing}

Right now, BaseSAFE uses the instrumentation of AFL QEMU mode.
For each new translation block, it calculates a shift and a XOR operation to find the \texttt{afl\_idx} offset, see Listing~\ref{lst:instrumentation}.

\vspace*{0.3cm}
\begin{minipage}{\linewidth}
\begin{lstlisting}[captionpos=b,frame=single,language=C,label=lst:instrumentation,caption=Instrumentation in BaseSAFE]
afl_idx = cur_loc ^ uc->afl_prev_loc
INC_AFL_AREA(afl_idx);
uc->afl_prev_loc = cur_loc >> 1
\end{lstlisting}
\end{minipage}

At the position of \texttt{afl\_idx}, \texttt{INC\_AFL\_AREA} then increases a counter at the shared map used to report feedback to \AFL{}.

While this hash is good enough to find almost all paths, our tests indicate that collisions occasionally occur.
This leaves a small number of branches undetected, negatively affecting feedback-based mutations.
The fuzzer will still eventually reach colliding paths, but may classify the edge as already taken, putting less weight on this test case.
In the future, BaseSAFE will be extended with collision-free instrumentation.
Initial tests without AFL's hashing indicate a higher total number of total paths found.

\subsection{Additional Targets}

In the course of this paper, we were merely able to highlight a small portion of a whole operating system.
The analyzed MediaTek firmware for the HTC One in question has 56 calls to \texttt{msg\_receive\_extq}, the IPC mechanism to receive messages from the system queue, alone.
As described in~\ref{sec:nucleus}, each of these calls is one specific task.
Each task may contain parsers for multiple different network packets and call one of them, depending on the ILM \texttt{msg\_id} (see Fig.~\ref{fig:structs} for one ILM example).
Each such queue read may contain, or depend on, user-provided input and can be an interesting target to fuzz in itself.
Especially 2G functions might be another interesting target as it is usually old code that may be seldomly used and tested.
Apart from MediaTek, other cellular basebands, base stations, and even unrelated firmware can be tested with BaseSAFE after some adaptation.
Piece by piece, support for non-standard architectures, such as Qualcomm's Hexagon, can be added by porting their existing open-source QEMU versions~\cite{qemuhexagon} to Unicorn engine, thereby improving the number of testable basebands.

\subsection{Further Harness Automation}

While the methods and tools presented in this paper will be applicable to all cellular basebands, hook creation could be further automated to allow adaption to new platforms with less manual work.
Of course, this will need a deeper automated understanding of unknown firmware blobs, using heuristics and automated static analysis.

\section{Conclusion}

BaseSAFE shows good results.
It provides fast fuzzing speeds, with around 15000 executions on a small server for our real-world use-case, emulating multiple Mediatek baseband parsers.
This proves that emulation is a good fit for automated bug discovery in baseband firmware analysis.
BaseSAFE's snapshot-based fuzzing introduces a new level of precision and achieves high coverage and execution speeds, not achievable by classical over-the-air fuzzing approaches.
BaseSAFE is built on top of state-of-the-art open-source tools and has---in turn---been open-sourced.
The API offered by its Rust bindings makes it easy to quickly implement fuzz cases without additional overhead.
We were able to run high-performance partial emulations of complex firmware.
The introduced drop-in sanitizing allocator finds memory corruption bugs automatically, solving a problem considered hard in prior literature~\cite{muench2018you}.
Our proposed framework found vulnerabilities that are reproducible over-the-air.
We conclude, that with BaseSAFE, fuzzing of embedded firmware in general, and different parts of the MediaTek cellular baseband in particular, is greatly facilitated and allows for efficient automated bug discovery in various scenarios.

\textbf{Responsible disclosure.} All results of this research were communicated to MediaTek in a timely fashion.

\subsection*{Acknowledgements}
The authors would like to thank Jiska Classen and Altaf Shaik for valuable feedback.

%% file: basesafe.bbl

\begin{thebibliography}{61}

%%% the Web address.

\ifx \showCODEN    \undefined \def \showCODEN     #1{\unskip}     \fi
\ifx \showDOI      \undefined \def \showDOI       #1{#1}\fi
\ifx \showISBNx    \undefined \def \showISBNx     #1{\unskip}     \fi
\ifx \showISBNxiii \undefined \def \showISBNxiii  #1{\unskip}     \fi
\ifx \showISSN     \undefined \def \showISSN      #1{\unskip}     \fi
\ifx \showLCCN     \undefined \def \showLCCN      #1{\unskip}     \fi
\ifx \shownote     \undefined \def \shownote      #1{#1}          \fi
\ifx \showarticletitle \undefined \def \showarticletitle #1{#1}   \fi
\ifx \showURL      \undefined \def \showURL       {\relax}        \fi
% invisible to TeX
\providecommand\bibfield[2]{#2}
\providecommand\bibinfo[2]{#2}
\providecommand\natexlab[1]{#1}
\providecommand\showeprint[2][]{arXiv:#2}

\bibitem[\protect\citeauthoryear{3GPP}{3GPP}{2018a}]%
        {3gpp.27.007}
\bibfield{author}{\bibinfo{person}{3GPP}.} \bibinfo{year}{2018}\natexlab{a}.
\newblock \bibinfo{booktitle}{\emph{{AT command set for User Equipment (UE)}}}.
\newblock \bibinfo{type}{Technical Specification (TS)} 27.007.
  \bibinfo{institution}{{3rd Generation Partnership Project (3GPP)}}.
\newblock
\urldef\tempurl%
\url{http://www.3gpp.org/DynaReport/27007.htm}
\showURL{%
\tempurl}
\newblock
\shownote{Version 15.4.0.}


\bibitem[\protect\citeauthoryear{3GPP}{3GPP}{2018b}]%
        {3gpp.23.040}
\bibfield{author}{\bibinfo{person}{3GPP}.} \bibinfo{year}{2018}\natexlab{b}.
\newblock \bibinfo{booktitle}{\emph{{Technical realization of the Short Message
  Service (SMS)}}}.
\newblock \bibinfo{type}{Technical Specification (TS)} 23.040.
  \bibinfo{institution}{{3rd Generation Partnership Project (3GPP)}}.
\newblock
\urldef\tempurl%
\url{http://www.3gpp.org/DynaReport/23040.htm}
\showURL{%
\tempurl}


\bibitem[\protect\citeauthoryear{3GPP}{3GPP}{2020a}]%
        {3gpp.36.331}
\bibfield{author}{\bibinfo{person}{3GPP}.} \bibinfo{year}{2020}\natexlab{a}.
\newblock \bibinfo{booktitle}{\emph{{Evolved Universal Terrestrial Radio Access
  (E-UTRA); Radio Resource Control (RRC); Protocol specification}}}.
\newblock \bibinfo{type}{Technical Specification (TS)} 36.331.
  \bibinfo{institution}{{3rd Generation Partnership Project (3GPP)}}.
\newblock
\urldef\tempurl%
\url{http://www.3gpp.org/DynaReport/36331.htm}
\showURL{%
\tempurl}


\bibitem[\protect\citeauthoryear{3GPP}{3GPP}{2020b}]%
        {3gpp.24.008}
\bibfield{author}{\bibinfo{person}{3GPP}.} \bibinfo{year}{2020}\natexlab{b}.
\newblock \bibinfo{booktitle}{\emph{{Mobile radio interface Layer 3
  specification; Core network protocols; Stage 3}}}.
\newblock \bibinfo{type}{Technical Specification (TS)} 24.008.
  \bibinfo{institution}{{3rd Generation Partnership Project (3GPP)}}.
\newblock
\urldef\tempurl%
\url{http://www.3gpp.org/DynaReport/24008.htm}
\showURL{%
\tempurl}


\bibitem[\protect\citeauthoryear{3GPP}{3GPP}{2020c}]%
        {3gpp.24.301}
\bibfield{author}{\bibinfo{person}{3GPP}.} \bibinfo{year}{2020}\natexlab{c}.
\newblock \bibinfo{booktitle}{\emph{{Non-Access-Stratum (NAS) protocol for
  Evolved Packet System (EPS); Stage 3}}}.
\newblock \bibinfo{type}{Technical Specification (TS)} 24.301.
  \bibinfo{institution}{{3rd Generation Partnership Project (3GPP)}}.
\newblock
\urldef\tempurl%
\url{http://www.3gpp.org/DynaReport/24301.htm}
\showURL{%
\tempurl}


\bibitem[\protect\citeauthoryear{3GPP}{3GPP}{2020d}]%
        {3gpp.25.331}
\bibfield{author}{\bibinfo{person}{3GPP}.} \bibinfo{year}{2020}\natexlab{d}.
\newblock \bibinfo{booktitle}{\emph{{Radio Resource Control (RRC); Protocol
  specification}}}.
\newblock \bibinfo{type}{Technical Specification (TS)} 25.331.
  \bibinfo{institution}{{3rd Generation Partnership Project (3GPP)}}.
\newblock
\urldef\tempurl%
\url{http://www.3gpp.org/DynaReport/25331.htm}
\showURL{%
\tempurl}


\bibitem[\protect\citeauthoryear{AFL}{AFL}{2020}]%
        {qemumode}
\bibfield{author}{\bibinfo{person}{AFL}.} \bibinfo{year}{2020}\natexlab{}.
\newblock \bibinfo{booktitle}{\emph{AFL QEMU Mode}}.
\newblock
\urldef\tempurl%
\url{https://github.com/mirrorer/afl/blob/master/qemu_mode/README.qemu}
\showURL{%
\tempurl}


\bibitem[\protect\citeauthoryear{Alecu}{Alecu}{2013}]%
        {Alecu2013}
\bibfield{author}{\bibinfo{person}{Bogdan Alecu}.}
  \bibinfo{year}{2013}\natexlab{}.
\newblock \showarticletitle{SMS fuzzing--SIM toolkit attack}.
\newblock \bibinfo{journal}{\emph{DEF CON 21}} (\bibinfo{year}{2013}).
\newblock


\bibitem[\protect\citeauthoryear{Bellard}{Bellard}{2005}]%
        {bellard2005qemu}
\bibfield{author}{\bibinfo{person}{Fabrice Bellard}.}
  \bibinfo{year}{2005}\natexlab{}.
\newblock \showarticletitle{QEMU, a fast and portable dynamic translator.}. In
  \bibinfo{booktitle}{\emph{USENIX Annual Technical Conference, FREENIX
  Track}}, Vol.~\bibinfo{volume}{41}. \bibinfo{pages}{46}.
\newblock


\bibitem[\protect\citeauthoryear{Bellard}{Bellard}{2020}]%
        {qemutcg}
\bibfield{author}{\bibinfo{person}{Fabrice Bellard}.}
  \bibinfo{year}{2020}\natexlab{}.
\newblock \bibinfo{booktitle}{\emph{Tiny Code Generator}}.
\newblock
\urldef\tempurl%
\url{https://git.qemu.org/?p=qemu.git;a=blob_plain;f=tcg/README;hb=HEAD}
\showURL{%
\tempurl}


\bibitem[\protect\citeauthoryear{Biondo}{Biondo}{2018}]%
        {improvingaflqemu}
\bibfield{author}{\bibinfo{person}{Andrea Biondo}.}
  \bibinfo{year}{2018}\natexlab{}.
\newblock \showarticletitle{Improving AFL's QEMU mode performance}.
\newblock \bibinfo{journal}{\emph{0x41414141 in ?? ()}} (\bibinfo{date}{Sep}
  \bibinfo{year}{2018}).
\newblock
\urldef\tempurl%
\url{https://abiondo.me/2018/09/21/improving-afl-qemu-mode}
\showURL{%
\tempurl}


\bibitem[\protect\citeauthoryear{Butti and Tinnes}{Butti and Tinnes}{2008}]%
        {butti2008discovering}
\bibfield{author}{\bibinfo{person}{Laurent Butti} {and} \bibinfo{person}{Julien
  Tinnes}.} \bibinfo{year}{2008}\natexlab{}.
\newblock \showarticletitle{Discovering and exploiting 802.11 wireless driver
  vulnerabilities}.
\newblock \bibinfo{journal}{\emph{Journal in Computer Virology}}
  \bibinfo{volume}{4}, \bibinfo{number}{1} (\bibinfo{year}{2008}),
  \bibinfo{pages}{25--37}.
\newblock


\bibitem[\protect\citeauthoryear{{Comsecuris}}{{Comsecuris}}{2020}]%
        {qemuhexagon}
\bibfield{author}{\bibinfo{person}{{Comsecuris}}.}
  \bibinfo{year}{2020}\natexlab{}.
\newblock \bibinfo{title}{{QEMU} with support for {QDSP6} user mode emulation}.
\newblock
\newblock
\urldef\tempurl%
\url{https://github.com/Comsecuris/qemu-hexagon}
\showURL{%
\tempurl}


\bibitem[\protect\citeauthoryear{Dinesh, Burow, Xu, and Payer}{Dinesh
  et~al\mbox{.}}{2020}]%
        {retrowrite}
\bibfield{author}{\bibinfo{person}{S.~Dinesh~S. Dinesh},
  \bibinfo{person}{Nathan Burow}, \bibinfo{person}{Dongyan Xu}, {and}
  \bibinfo{person}{Mathias Payer}.} \bibinfo{year}{2020}\natexlab{}.
\newblock \showarticletitle{RetroWrite: Statically Instrumenting COTS Binaries
  for Fuzzing and Sanitization}. In \bibinfo{booktitle}{\emph{IEEE S\&P 2020}}.
\newblock


\bibitem[\protect\citeauthoryear{Drysdale}{Drysdale}{2016}]%
        {syzkaller}
\bibfield{author}{\bibinfo{person}{David Drysdale}.}
  \bibinfo{year}{2016}\natexlab{}.
\newblock \bibinfo{booktitle}{\emph{Coverage-guided kernel fuzzing with
  syzkaller}}.
\newblock
\urldef\tempurl%
\url{https://lwn.net/Articles/677764/}
\showURL{%
\tempurl}


\bibitem[\protect\citeauthoryear{Duquette}{Duquette}{2020}]%
        {unicorn-rs}
\bibfield{author}{\bibinfo{person}{Sébastien Duquette}.}
  \bibinfo{year}{2020}\natexlab{}.
\newblock \bibinfo{booktitle}{\emph{Rust bindings for the unicorn {CPU}
  emulator}}.
\newblock
\urldef\tempurl%
\url{https://github.com/ekse/unicorn-rs}
\showURL{%
\tempurl}


\bibitem[\protect\citeauthoryear{Falk}{Falk}{2018}]%
        {vectoremu}
\bibfield{author}{\bibinfo{person}{Brandon Falk}.}
  \bibinfo{year}{2018}\natexlab{}.
\newblock \bibinfo{title}{{Vectorized Emulation: Hardware accelerated taint
  tracking at 2 trillion instructions per second}}.
\newblock
\newblock
\urldef\tempurl%
\url{https://gamozolabs.github.io/fuzzing/2018/10/14/vectorized_emulation.html}
\showURL{%
\tempurl}
\newblock
\shownote{[Online; accessed 11. Nov. 2018].}


\bibitem[\protect\citeauthoryear{Fioraldi}{Fioraldi}{2019}]%
        {qasan}
\bibfield{author}{\bibinfo{person}{Andrea Fioraldi}.}
  \bibinfo{year}{2019}\natexlab{}.
\newblock \bibinfo{title}{Sanitized Emulation with {QASan}}.
\newblock
\newblock
\urldef\tempurl%
\url{https://andreafioraldi.github.io/articles/2019/12/20/sanitized-emulation-with-qasan.html}
\showURL{%
\tempurl}


\bibitem[\protect\citeauthoryear{Golde and Komaromy}{Golde and
  Komaromy}{2016}]%
        {Golde2016}
\bibfield{author}{\bibinfo{person}{Nico Golde} {and} \bibinfo{person}{Daniel
  Komaromy}.} \bibinfo{year}{2016}\natexlab{}.
\newblock \bibinfo{booktitle}{\emph{Breaking Band: reverse engineering and
  exploiting the shannon baseband}}.
\newblock
\urldef\tempurl%
\url{https://comsecuris.com/slides/recon2016-breaking_band.pdf}
\showURL{%
\tempurl}


\bibitem[\protect\citeauthoryear{Gomez{-}Miguelez, Garcia{-}Saavedra, Sutton,
  Serrano, Cano, and Leith}{Gomez{-}Miguelez et~al\mbox{.}}{2016}]%
        {srslte}
\bibfield{author}{\bibinfo{person}{Ismael Gomez{-}Miguelez},
  \bibinfo{person}{Andres Garcia{-}Saavedra}, \bibinfo{person}{Paul~D. Sutton},
  \bibinfo{person}{Pablo Serrano}, \bibinfo{person}{Cristina Cano}, {and}
  \bibinfo{person}{Douglas~J. Leith}.} \bibinfo{year}{2016}\natexlab{}.
\newblock \showarticletitle{srsLTE: An Open-Source Platform for {LTE} Evolution
  and Experimentation}.
\newblock \bibinfo{journal}{\emph{CoRR}}  \bibinfo{volume}{abs/1602.04629}
  (\bibinfo{year}{2016}).
\newblock
\urldef\tempurl%
\url{http://arxiv.org/abs/1602.04629}
\showURL{%
\tempurl}


\bibitem[\protect\citeauthoryear{Grassi and Chen}{Grassi and Chen}{2020}]%
        {grassi}
\bibfield{author}{\bibinfo{person}{Marco Grassi} {and} \bibinfo{person}{Xingyu
  Chen}.} \bibinfo{year}{2020}\natexlab{}.
\newblock \showarticletitle{Exploring the {MediaTek} Baseband}. In
  \bibinfo{booktitle}{\emph{OffensiveCon}}.
\newblock


\bibitem[\protect\citeauthoryear{Hay}{Hay}{2017}]%
        {Hay2017}
\bibfield{author}{\bibinfo{person}{Roee Hay}.} \bibinfo{year}{2017}\natexlab{}.
\newblock \showarticletitle{fastboot {OEM} vuln: Android bootloader
  vulnerabilities in vendor customizations}. In \bibinfo{booktitle}{\emph{11th
  {USENIX} Workshop on Offensive Technologies ({WOOT} 17)}}.
\newblock


\bibitem[\protect\citeauthoryear{Hengeveld}{Hengeveld}{2013}]%
        {gsmkhexagon}
\bibfield{author}{\bibinfo{person}{Willem Hengeveld}.}
  \bibinfo{year}{2013}\natexlab{}.
\newblock \bibinfo{booktitle}{\emph{IDA processor module for the hexagon
  (QDSP6) processor}}.
\newblock
\urldef\tempurl%
\url{https://github.com/gsmk/hexagon}
\showURL{%
\tempurl}


\bibitem[\protect\citeauthoryear{Hernandez and Butler}{Hernandez and
  Butler}{2019}]%
        {Hernandez2019}
\bibfield{author}{\bibinfo{person}{Grant Hernandez} {and}
  \bibinfo{person}{Kevin R.~B. Butler}.} \bibinfo{year}{2019}\natexlab{}.
\newblock \showarticletitle{Basebads: Automated Security Analysis of Baseband
  Firmware: Poster}.
\newblock  (\bibinfo{year}{2019}), \bibinfo{pages}{318–319}.
\newblock
\showISBNx{9781450367264}
\urldef\tempurl%
\url{https://doi.org/10.1145/3317549.3326310}
\showDOI{\tempurl}


\bibitem[\protect\citeauthoryear{Hertz and Newsham}{Hertz and Newsham}{2016}]%
        {triforce}
\bibfield{author}{\bibinfo{person}{J Hertz} {and} \bibinfo{person}{T Newsham}.}
  \bibinfo{year}{2016}\natexlab{}.
\newblock \showarticletitle{Project triforce: Run afl on everything}.
\newblock \bibinfo{journal}{\emph{NCC Group, Tech. Rep.}}
  (\bibinfo{year}{2016}).
\newblock


\bibitem[\protect\citeauthoryear{Heuse, Ei{\ss}feld, Fioraldi, and Maier}{Heuse
  et~al\mbox{.}}{2020}]%
        {aflplusplus}
\bibfield{author}{\bibinfo{person}{Marc Heuse}, \bibinfo{person}{Heiko
  Ei{\ss}feld}, \bibinfo{person}{Andrea Fioraldi}, {and}
  \bibinfo{person}{Dominik Maier}.} \bibinfo{year}{2020}\natexlab{}.
\newblock \bibinfo{title}{american fuzzy lop plus plus (afl++)}.
\newblock \bibinfo{howpublished}{GitHub}.
\newblock
\urldef\tempurl%
\url{https://github.com/vanhauser-thc/AFLplusplus}
\showURL{%
\tempurl}


\bibitem[\protect\citeauthoryear{Hussain, Chowdhury, Mehnaz, and
  Bertino}{Hussain et~al\mbox{.}}{2018}]%
        {Hussain2018}
\bibfield{author}{\bibinfo{person}{Syed Hussain}, \bibinfo{person}{Omar
  Chowdhury}, \bibinfo{person}{Shagufta Mehnaz}, {and} \bibinfo{person}{Elisa
  Bertino}.} \bibinfo{year}{2018}\natexlab{}.
\newblock \showarticletitle{LTEInspector: A systematic approach for adversarial
  testing of 4G LTE}. In \bibinfo{booktitle}{\emph{Network and Distributed
  Systems Security (NDSS) Symposium 2018}}.
\newblock


\bibitem[\protect\citeauthoryear{Hussain, Echeverria, Karim, Chowdhury, and
  Bertino}{Hussain et~al\mbox{.}}{2019}]%
        {Hussain2019}
\bibfield{author}{\bibinfo{person}{Syed~Rafiul Hussain},
  \bibinfo{person}{Mitziu Echeverria}, \bibinfo{person}{Imtiaz Karim},
  \bibinfo{person}{Omar Chowdhury}, {and} \bibinfo{person}{Elisa Bertino}.}
  \bibinfo{year}{2019}\natexlab{}.
\newblock \showarticletitle{5GReasoner: {A} Property-Directed Security and
  Privacy Analysis Framework for 5G Cellular Network Protocol}. In
  \bibinfo{booktitle}{\emph{Proceedings of the 2019 {ACM} {SIGSAC} Conference
  on Computer and Communications Security, {CCS} 2019, London, UK, November
  11-15, 2019}}, \bibfield{editor}{\bibinfo{person}{Lorenzo Cavallaro},
  \bibinfo{person}{Johannes Kinder}, \bibinfo{person}{XiaoFeng Wang}, {and}
  \bibinfo{person}{Jonathan Katz}} (Eds.). \bibinfo{publisher}{{ACM}},
  \bibinfo{pages}{669--684}.
\newblock
\urldef\tempurl%
\url{https://doi.org/10.1145/3319535.3354263}
\showDOI{\tempurl}


\bibitem[\protect\citeauthoryear{{Imagination Technologies}}{{Imagination
  Technologies}}{2017}]%
        {mipsmtk}
\bibfield{author}{\bibinfo{person}{{Imagination Technologies}}.}
  \bibinfo{year}{2017}\natexlab{}.
\newblock \bibinfo{booktitle}{\emph{{MediaTek} selects {MIPS} for {LTE}
  modems}}.
\newblock
\urldef\tempurl%
\url{https://www.mips.com/press/mediatek-selects-mips-for-lte-modems/}
\showURL{%
\tempurl}


\bibitem[\protect\citeauthoryear{{Johansson}, {Svensson}, {Larson}, {Almgren},
  and {Gulisano}}{{Johansson} et~al\mbox{.}}{2014}]%
        {Johansson2014}
\bibfield{author}{\bibinfo{person}{W. {Johansson}}, \bibinfo{person}{M.
  {Svensson}}, \bibinfo{person}{U.~E. {Larson}}, \bibinfo{person}{M.
  {Almgren}}, {and} \bibinfo{person}{V. {Gulisano}}.}
  \bibinfo{year}{2014}\natexlab{}.
\newblock \showarticletitle{T-Fuzz: Model-Based Fuzzing for Robustness Testing
  of Telecommunication Protocols}. In \bibinfo{booktitle}{\emph{2014 IEEE
  Seventh International Conference on Software Testing, Verification and
  Validation}}. \bibinfo{pages}{323--332}.
\newblock
\showISSN{2159-4848}
\urldef\tempurl%
\url{https://doi.org/10.1109/ICST.2014.45}
\showDOI{\tempurl}


\bibitem[\protect\citeauthoryear{Karim, Cicala, Hussain, Chowdhury, and
  Bertino}{Karim et~al\mbox{.}}{2019}]%
        {Karim2019}
\bibfield{author}{\bibinfo{person}{Imtiaz Karim}, \bibinfo{person}{Fabrizio
  Cicala}, \bibinfo{person}{Syed Hussain}, \bibinfo{person}{Omar Chowdhury},
  {and} \bibinfo{person}{Elisa Bertino}.} \bibinfo{year}{2019}\natexlab{}.
\newblock \showarticletitle{Opening Pandora's box through ATFuzzer: dynamic
  analysis of AT interface for Android smartphones}. \bibinfo{pages}{529--543}.
\newblock
\showISBNx{978-1-4503-7628-0}
\urldef\tempurl%
\url{https://doi.org/10.1145/3359789.3359833}
\showDOI{\tempurl}


\bibitem[\protect\citeauthoryear{Kim, Lee, Eunkyu, and Kim}{Kim
  et~al\mbox{.}}{2019}]%
        {Kim2019}
\bibfield{author}{\bibinfo{person}{Hongil Kim}, \bibinfo{person}{Jiho Lee},
  \bibinfo{person}{Lee Eunkyu}, {and} \bibinfo{person}{Yongdae Kim}.}
  \bibinfo{year}{2019}\natexlab{}.
\newblock \showarticletitle{Touching the Untouchables: Dynamic Security
  Analysis of the LTE Control Plane}. In \bibinfo{booktitle}{\emph{2019 IEEE
  Symposium on Security and Privacy (SP)}}. \bibinfo{pages}{1153--1168}.
\newblock
\showISSN{1081-6011}
\urldef\tempurl%
\url{https://doi.org/10.1109/SP.2019.00038}
\showDOI{\tempurl}


\bibitem[\protect\citeauthoryear{{Lu}, {He}, and {Li}}{{Lu}
  et~al\mbox{.}}{2011}]%
        {freetibet}
\bibfield{author}{\bibinfo{person}{X. {Lu}}, \bibinfo{person}{J. {He}}, {and}
  \bibinfo{person}{J. {Li}}.} \bibinfo{year}{2011}\natexlab{}.
\newblock \showarticletitle{A Tibetan input method based on MTK for mobile
  phone}. In \bibinfo{booktitle}{\emph{2011 International Conference on
  Consumer Electronics, Communications and Networks (CECNet)}}.
  \bibinfo{pages}{3884--3887}.
\newblock
\showISSN{null}
\urldef\tempurl%
\url{https://doi.org/10.1109/CECNET.2011.5768296}
\showDOI{\tempurl}


\bibitem[\protect\citeauthoryear{Maier, Radtke, and Harren}{Maier
  et~al\mbox{.}}{2019}]%
        {ucf}
\bibfield{author}{\bibinfo{person}{Dominik Maier}, \bibinfo{person}{Benedikt
  Radtke}, {and} \bibinfo{person}{Bastian Harren}.}
  \bibinfo{year}{2019}\natexlab{}.
\newblock \showarticletitle{Unicorefuzz: On the Viability of Emulation for
  Kernelspace Fuzzing}. In \bibinfo{booktitle}{\emph{13th {USENIX} Workshop on
  Offensive Technologies, {WOOT} 2019, Santa Clara, CA, USA, August 12-13,
  2019}}, \bibfield{editor}{\bibinfo{person}{Alex Gantman} {and}
  \bibinfo{person}{Cl{\'{e}}mentine Maurice}} (Eds.).
  \bibinfo{publisher}{{USENIX} Association}.
\newblock
\urldef\tempurl%
\url{https://www.usenix.org/conference/woot19/presentation/maier}
\showURL{%
\tempurl}


\bibitem[\protect\citeauthoryear{Miru}{Miru}{2017}]%
        {Miru2017}
\bibfield{author}{\bibinfo{person}{György Miru}.}
  \bibinfo{year}{2017}\natexlab{}.
\newblock \bibinfo{booktitle}{\emph{Path of Least Resistance: Cellular Baseband
  to Application Processor Escalation on Mediatek Devices}}.
\newblock
\urldef\tempurl%
\url{https://comsecuris.com/blog/posts/path_of_least_resistance/}
\showURL{%
\tempurl}


\bibitem[\protect\citeauthoryear{{M}uench, {F}rancillon, and
  {B}alzarotti}{{M}uench et~al\mbox{.}}{2018}]%
        {Avatar2}
\bibfield{author}{\bibinfo{person}{{M}arius {M}uench},
  \bibinfo{person}{{A}ur{\'e}lien {F}rancillon}, {and}
  \bibinfo{person}{{D}avide {B}alzarotti}.} \bibinfo{year}{2018}\natexlab{}.
\newblock \showarticletitle{{A}vatar²: {A} multi-target orchestration
  platform}. In \bibinfo{booktitle}{\emph{{BAR} 2018, {W}orkshop on {B}inary
  {A}nalysis {R}esearch, colocated with {NDSS} {S}ymposium, 18 {F}ebruary 2018,
  {S}an {D}iego, {USA}}}. \bibinfo{address}{{S}an {D}iego, {UNITED} {STATES}}.
\newblock
\urldef\tempurl%
\url{http://www.eurecom.fr/publication/5437}
\showURL{%
\tempurl}


\bibitem[\protect\citeauthoryear{Muench, Stijohann, Kargl, Francillon, and
  Balzarotti}{Muench et~al\mbox{.}}{2018}]%
        {muench2018you}
\bibfield{author}{\bibinfo{person}{Marius Muench}, \bibinfo{person}{Jan
  Stijohann}, \bibinfo{person}{Frank Kargl}, \bibinfo{person}{Aur{\'e}lien
  Francillon}, {and} \bibinfo{person}{Davide Balzarotti}.}
  \bibinfo{year}{2018}\natexlab{}.
\newblock \showarticletitle{What you corrupt is not what you crash: Challenges
  in fuzzing embedded devices}. In \bibinfo{booktitle}{\emph{Proceedings 2018
  Network and Distributed System Security Symposium, San Diego, CA}}.
\newblock


\bibitem[\protect\citeauthoryear{Mulliner, Borgaonkar, Stewin, and
  Seifert}{Mulliner et~al\mbox{.}}{2013}]%
        {Mulliner2013}
\bibfield{author}{\bibinfo{person}{Collin Mulliner},
  \bibinfo{person}{Ravishankar Borgaonkar}, \bibinfo{person}{Patrick Stewin},
  {and} \bibinfo{person}{Jean-Pierre Seifert}.}
  \bibinfo{year}{2013}\natexlab{}.
\newblock \showarticletitle{SMS-Based One-Time Passwords: Attacks and Defense}.
  In \bibinfo{booktitle}{\emph{Detection of Intrusions and Malware, and
  Vulnerability Assessment}}, \bibfield{editor}{\bibinfo{person}{Konrad Rieck},
  \bibinfo{person}{Patrick Stewin}, {and} \bibinfo{person}{Jean-Pierre
  Seifert}} (Eds.). \bibinfo{publisher}{Springer Berlin Heidelberg},
  \bibinfo{address}{Berlin, Heidelberg}, \bibinfo{pages}{150--159}.
\newblock
\showISBNx{978-3-642-39235-1}


\bibitem[\protect\citeauthoryear{Mulliner, Golde, and Seifert}{Mulliner
  et~al\mbox{.}}{2011}]%
        {Mulliner2011}
\bibfield{author}{\bibinfo{person}{Collin Mulliner}, \bibinfo{person}{Nico
  Golde}, {and} \bibinfo{person}{Jean-Pierre Seifert}.}
  \bibinfo{year}{2011}\natexlab{}.
\newblock \showarticletitle{{SMS of Death: from analyzing to attacking mobile
  phones on a large scale}}.
\newblock \bibinfo{journal}{\emph{USENIX Security}} (\bibinfo{year}{2011}).
\newblock
\urldef\tempurl%
\url{http://static.usenix.org/events/sec11/tech/full{\_}papers/Mulliner.pdf}
\showURL{%
\tempurl}


\bibitem[\protect\citeauthoryear{Mulliner and Miller}{Mulliner and
  Miller}{2009}]%
        {Mulliner2009}
\bibfield{author}{\bibinfo{person}{Collin Mulliner} {and}
  \bibinfo{person}{Charlie Miller}.} \bibinfo{year}{2009}\natexlab{}.
\newblock \showarticletitle{Fuzzing the Phone in your Phone}.
\newblock \bibinfo{journal}{\emph{Black Hat USA 2009}} (\bibinfo{year}{2009}).
\newblock
\urldef\tempurl%
\url{https://www.blackhat.com/presentations/bh-usa-09/MILLER/BHUSA09-Miller-FuzzingPhone-PAPER.pdf}
\showURL{%
\tempurl}


\bibitem[\protect\citeauthoryear{Ngyuen and Dang}{Ngyuen and Dang}{2020}]%
        {unicornemu}
\bibfield{author}{\bibinfo{person}{Anh~Quynh Ngyuen} {and}
  \bibinfo{person}{Hoang~Vu Dang}.} \bibinfo{year}{2020}\natexlab{}.
\newblock \bibinfo{booktitle}{\emph{Unicorn: Next Generation CPU Emulator
  Framework}}.
\newblock
\urldef\tempurl%
\url{http://www.unicorn-engine.org/BHUSA2015-unicorn.pdf}
\showURL{%
\tempurl}


\bibitem[\protect\citeauthoryear{{OpenBTS}}{{OpenBTS}}{2020}]%
        {openbts-umts}
\bibfield{author}{\bibinfo{person}{{OpenBTS}}.}
  \bibinfo{year}{2020}\natexlab{}.
\newblock \bibinfo{title}{{OpenBTS-UMTS}}.
\newblock
\newblock
\urldef\tempurl%
\url{http://openbts.org/w/index.php?title=OpenBTS-UMTS}
\showURL{%
\tempurl}


\bibitem[\protect\citeauthoryear{{Osmocom Project}}{{Osmocom Project}}{2020}]%
        {osmocom}
\bibfield{author}{\bibinfo{person}{{Osmocom Project}}.}
  \bibinfo{year}{2020}\natexlab{}.
\newblock \bibinfo{title}{{Cellular Network Infrastructure}}.
\newblock
\newblock
\urldef\tempurl%
\url{https://osmocom.org/projects/cellular-infrastructure/wiki}
\showURL{%
\tempurl}


\bibitem[\protect\citeauthoryear{{P1 Security}}{{P1 Security}}{2020}]%
        {p1tf}
\bibfield{author}{\bibinfo{person}{{P1 Security}}.}
  \bibinfo{year}{2020}\natexlab{}.
\newblock \bibinfo{booktitle}{\emph{P1 Telecom Fuzzer}}.
\newblock
\urldef\tempurl%
\url{https://www.p1sec.com/corp/products/p1-telecom-fuzzer-ptf/}
\showURL{%
\tempurl}


\bibitem[\protect\citeauthoryear{Park}{Park}{2017}]%
        {scat}
\bibfield{author}{\bibinfo{person}{Shinjo Park}.}
  \bibinfo{year}{2017}\natexlab{}.
\newblock \bibinfo{booktitle}{\emph{SCAT: Signaling Collection and Analysis
  Tool}}.
\newblock
\urldef\tempurl%
\url{https://github.com/fgsect/scat}
\showURL{%
\tempurl}


\bibitem[\protect\citeauthoryear{Park, Shaik, Borgaonkar, and Seifert}{Park
  et~al\mbox{.}}{2016}]%
        {Park2016}
\bibfield{author}{\bibinfo{person}{Shinjo Park}, \bibinfo{person}{Altaf Shaik},
  \bibinfo{person}{Ravishankar Borgaonkar}, {and} \bibinfo{person}{Jean-Pierre
  Seifert}.} \bibinfo{year}{2016}\natexlab{}.
\newblock \showarticletitle{White Rabbit in Mobile: Effect of Unsecured Clock
  Source in Smartphones}. In \bibinfo{booktitle}{\emph{Proceedings of the 6th
  Workshop on Security and Privacy in Smartphones and Mobile Devices}}. ACM,
  \bibinfo{pages}{13--21}.
\newblock


\bibitem[\protect\citeauthoryear{Rupprecht, Dabrowski, Holz, Weippl, and
  P{\"{o}}pper}{Rupprecht et~al\mbox{.}}{2017}]%
        {Rupprecht2017}
\bibfield{author}{\bibinfo{person}{David Rupprecht}, \bibinfo{person}{Adrian
  Dabrowski}, \bibinfo{person}{Thorsten Holz}, \bibinfo{person}{Edgar Weippl},
  {and} \bibinfo{person}{Christina P{\"{o}}pper}.}
  \bibinfo{year}{2017}\natexlab{}.
\newblock \showarticletitle{{On Security Research Towards Future Mobile Network
  Generations}}.
\newblock  (\bibinfo{date}{oct} \bibinfo{year}{2017}).
\newblock
\showeprint[arxiv]{1710.08932}
\urldef\tempurl%
\url{http://arxiv.org/abs/1710.08932}
\showURL{%
\tempurl}


\bibitem[\protect\citeauthoryear{Rupprecht, Jansen, and P{\"o}pper}{Rupprecht
  et~al\mbox{.}}{2016}]%
        {Rupprecht2016}
\bibfield{author}{\bibinfo{person}{David Rupprecht}, \bibinfo{person}{Kai
  Jansen}, {and} \bibinfo{person}{Christina P{\"o}pper}.}
  \bibinfo{year}{2016}\natexlab{}.
\newblock \showarticletitle{Putting {LTE} Security Functions to the Test: A
  Framework to Evaluate Implementation Correctness}. In
  \bibinfo{booktitle}{\emph{10th {USENIX} Workshop on Offensive Technologies
  ({WOOT} 16)}}.
\newblock


\bibitem[\protect\citeauthoryear{Rupprecht, Kohls, Holz, and
  P{\"{o}}pper}{Rupprecht et~al\mbox{.}}{2019}]%
        {Rupprecht2019}
\bibfield{author}{\bibinfo{person}{David Rupprecht}, \bibinfo{person}{Katharina
  Kohls}, \bibinfo{person}{Thorsten Holz}, {and} \bibinfo{person}{Christina
  P{\"{o}}pper}.} \bibinfo{year}{2019}\natexlab{}.
\newblock \showarticletitle{{Breaking LTE on Layer Two}}. In
  \bibinfo{booktitle}{\emph{2019 IEEE Symposium on Security and Privacy (SP)}}.
\newblock


\bibitem[\protect\citeauthoryear{Schumilo, Aschermann, Abbasi, Worner, and
  Holz}{Schumilo et~al\mbox{.}}{2020}]%
        {hypercube}
\bibfield{author}{\bibinfo{person}{Sergej Schumilo}, \bibinfo{person}{Cornelius
  Aschermann}, \bibinfo{person}{Ali Abbasi}, \bibinfo{person}{Simon Worner},
  {and} \bibinfo{person}{Thorsten Holz}.} \bibinfo{year}{2020}\natexlab{}.
\newblock \showarticletitle{{HYPER-CUBE}: High-Dimensional Hypervisor Fuzzing}.
  In \bibinfo{booktitle}{\emph{27th Annual Network and Distributed System
  Security Symposium, {NDSS} 2017, San Diego, California, USA, 2020}}.
\newblock
\urldef\tempurl%
\url{https://doi.org/10.14722/ndss.2020.23096}
\showDOI{\tempurl}


\bibitem[\protect\citeauthoryear{Schumilo, Aschermann, Gawlik, Schinzel, and
  Holz}{Schumilo et~al\mbox{.}}{2017}]%
        {kAFL}
\bibfield{author}{\bibinfo{person}{Sergej Schumilo}, \bibinfo{person}{Cornelius
  Aschermann}, \bibinfo{person}{Robert Gawlik}, \bibinfo{person}{Sebastian
  Schinzel}, {and} \bibinfo{person}{Thorsten Holz}.}
  \bibinfo{year}{2017}\natexlab{}.
\newblock \showarticletitle{kAFL: Hardware-Assisted Feedback Fuzzing for {OS}
  Kernels}. In \bibinfo{booktitle}{\emph{26th {USENIX} Security Symposium
  ({USENIX} Security 17)}}. \bibinfo{publisher}{{USENIX} Association},
  \bibinfo{address}{Vancouver, BC}, \bibinfo{pages}{167--182}.
\newblock
\showISBNx{978-1-931971-40-9}


\bibitem[\protect\citeauthoryear{Shaik, Borgaonkar, Asokan, Niemi, and
  Seifert}{Shaik et~al\mbox{.}}{2015}]%
        {Shaik2015}
\bibfield{author}{\bibinfo{person}{Altaf Shaik}, \bibinfo{person}{Ravishankar
  Borgaonkar}, \bibinfo{person}{N. Asokan}, \bibinfo{person}{Valtteri Niemi},
  {and} \bibinfo{person}{Jean-Pierre Seifert}.}
  \bibinfo{year}{2015}\natexlab{}.
\newblock \showarticletitle{{Practical attacks against privacy and availability
  in 4G/LTE mobile communication systems}}.
\newblock  (\bibinfo{year}{2015}).
\newblock
\urldef\tempurl%
\url{http://arxiv.org/abs/1510.07563}
\showURL{%
\tempurl}


\bibitem[\protect\citeauthoryear{Song, Hetzelt, Das, Spensky, Na, Volckaert,
  Vigna, Kruegel, Seifert, and Franz}{Song et~al\mbox{.}}{2019}]%
        {periscope}
\bibfield{author}{\bibinfo{person}{Dokyung Song}, \bibinfo{person}{Felicitas
  Hetzelt}, \bibinfo{person}{Dipanjan Das}, \bibinfo{person}{Chad Spensky},
  \bibinfo{person}{Yeoul Na}, \bibinfo{person}{Stijn Volckaert},
  \bibinfo{person}{Giovanni Vigna}, \bibinfo{person}{Christopher Kruegel},
  \bibinfo{person}{Jean{-}Pierre Seifert}, {and} \bibinfo{person}{Michael
  Franz}.} \bibinfo{year}{2019}\natexlab{}.
\newblock \showarticletitle{PeriScope: An Effective Probing and Fuzzing
  Framework for the Hardware-OS Boundary}. In \bibinfo{booktitle}{\emph{26th
  Annual Network and Distributed System Security Symposium, {NDSS} 2019, San
  Diego, California, USA, February 24-27, 2019}}. \bibinfo{publisher}{The
  Internet Society}.
\newblock
\urldef\tempurl%
\url{https://www.ndss-symposium.org/ndss-paper/periscope-an-effective-probing-and-fuzzing-framework-for-the-hardware-os-boundary/}
\showURL{%
\tempurl}


\bibitem[\protect\citeauthoryear{Strobel}{Strobel}{2007}]%
        {Strobel2007}
\bibfield{author}{\bibinfo{person}{Daehyun Strobel}.}
  \bibinfo{year}{2007}\natexlab{}.
\newblock \showarticletitle{{IMSI Catcher}}.
\newblock \bibinfo{journal}{\emph{Chair for Communication Security,
  Ruhr-Universit{\"a}t Bochum}} (\bibinfo{year}{2007}).
\newblock


\bibitem[\protect\citeauthoryear{Tian, Hernandez, Choi, Frost, Raules, Traynor,
  Vijayakumar, Harrison, Rahmati, Grace, et~al\mbox{.}}{Tian
  et~al\mbox{.}}{2018}]%
        {Tian2018}
\bibfield{author}{\bibinfo{person}{Dave~Jing Tian}, \bibinfo{person}{Grant
  Hernandez}, \bibinfo{person}{Joseph~I Choi}, \bibinfo{person}{Vanessa Frost},
  \bibinfo{person}{Christie Raules}, \bibinfo{person}{Patrick Traynor},
  \bibinfo{person}{Hayawardh Vijayakumar}, \bibinfo{person}{Lee Harrison},
  \bibinfo{person}{Amir Rahmati}, \bibinfo{person}{Michael Grace},
  {et~al\mbox{.}}} \bibinfo{year}{2018}\natexlab{}.
\newblock \showarticletitle{ATtention Spanned: Comprehensive Vulnerability
  Analysis of $\{$AT$\}$ Commands Within the Android Ecosystem}. In
  \bibinfo{booktitle}{\emph{27th {USENIX} Security Symposium ({USENIX} Security
  18)}}. \bibinfo{pages}{273--290}.
\newblock


\bibitem[\protect\citeauthoryear{Voss}{Voss}{2017}]%
        {aflunicorn}
\bibfield{author}{\bibinfo{person}{Nathan Voss}.}
  \bibinfo{year}{2017}\natexlab{}.
\newblock \bibinfo{booktitle}{\emph{afl-unicorn: Fuzzing Arbitrary Binary
  Code}}.
\newblock
\urldef\tempurl%
\url{https://hackernoon.com/afl-unicorn-fuzzing-arbitrary-binary-code-563ca28936bf}
\showURL{%
\tempurl}


\bibitem[\protect\citeauthoryear{Weinmann}{Weinmann}{2012}]%
        {Weinmann2012}
\bibfield{author}{\bibinfo{person}{Ralf-Philipp Weinmann}.}
  \bibinfo{year}{2012}\natexlab{}.
\newblock \showarticletitle{{Baseband Attacks: Remote Exploitation of Memory
  Corruptions in Cellular Protocol Stacks}}.
\newblock \bibinfo{journal}{\emph{USENIX Workshop on Offensive Technologies}}
  (\bibinfo{year}{2012}).
\newblock


\bibitem[\protect\citeauthoryear{{WikiChip}}{{WikiChip}}{2020}]%
        {wikichipx10}
\bibfield{author}{\bibinfo{person}{{WikiChip}}.}
  \bibinfo{year}{2020}\natexlab{}.
\newblock \bibinfo{booktitle}{\emph{{Helio X10 (MT6795) - MediaTek}}}.
\newblock
\urldef\tempurl%
\url{https://en.wikichip.org/wiki/mediatek/helio/mt6795}
\showURL{%
\tempurl}


\bibitem[\protect\citeauthoryear{Wojtowicz}{Wojtowicz}{[n.d.]}]%
        {openlte}
\bibfield{author}{\bibinfo{person}{Ben Wojtowicz}.}
  \bibinfo{year}{[n.d.]}\natexlab{}.
\newblock \bibinfo{title}{{OpenLTE}}.
\newblock
\newblock
\urldef\tempurl%
\url{http://openlte.sourceforge.net/}
\showURL{%
\tempurl}


\bibitem[\protect\citeauthoryear{Zalewski}{Zalewski}{2016}]%
        {zalewski2016technical}
\bibfield{author}{\bibinfo{person}{Michael Zalewski}.}
  \bibinfo{year}{2016}\natexlab{}.
\newblock \bibinfo{title}{Technical "whitepaper" for {AFL}-fuzz}.
\newblock
\newblock
\urldef\tempurl%
\url{http://lcamtuf.coredump.cx/afl/}
\showURL{%
\tempurl}


\bibitem[\protect\citeauthoryear{Zhang, Zhou, Luo, Wu, and Min}{Zhang
  et~al\mbox{.}}{2018}]%
        {fasterAFLPT}
\bibfield{author}{\bibinfo{person}{G. Zhang}, \bibinfo{person}{X. Zhou},
  \bibinfo{person}{Y. Luo}, \bibinfo{person}{X. Wu}, {and} \bibinfo{person}{E.
  Min}.} \bibinfo{year}{2018}\natexlab{}.
\newblock \showarticletitle{PTfuzz: Guided Fuzzing With Processor Trace
  Feedback}.
\newblock \bibinfo{journal}{\emph{IEEE Access}}  \bibinfo{volume}{6}
  (\bibinfo{year}{2018}), \bibinfo{pages}{37302--37313}.
\newblock
\showISSN{2169-3536}
\urldef\tempurl%
\url{https://doi.org/10.1109/ACCESS.2018.2851237}
\showDOI{\tempurl}


\end{thebibliography}
